\documentclass[11pt]{article}
\usepackage{ascmac}

\usepackage{amsmath, amssymb, amsfonts}
\usepackage[pdftex]{graphicx}
\usepackage{theorem}
\usepackage{here}
\usepackage{booktabs}
\usepackage[english]{babel}
\usepackage{natbib}
\usepackage{bm}
\usepackage{url}
\usepackage[utf8]{inputenc}

\theorembodyfont{\rmfamily}

\usepackage{soul}

\def\be{{\beta}}

\def\de{{\delta}}
\def\ep{{\varepsilon}}

\def\th{{\theta}}
\def\ka{{\kappa}}
\def\la{{\lambda}}
\def\si{{\sigma}}
\def\ta{{\tau}}

\def\a{{\text{\boldmath $a$}}}
\def\b{{\text{\boldmath $b$}}}

\def\f{{\text{\boldmath $f$}}}

\def\u{{\text{\boldmath $u$}}}
\def\v{{\text{\boldmath $v$}}}

\def\x{{\text{\boldmath $x$}}}
\def\y{{\text{\boldmath $y$}}}
\def\z{{\text{\boldmath $z$}}}

\def\D{{\text{\boldmath $D$}}}

\def\V{{\text{\boldmath $V$}}}

\def\X{{\text{\boldmath $X$}}}

\def\Z{{\text{\boldmath $Z$}}}

\def\bbe{{\text{\boldmath $\beta$}}}
\def\bga{{\text{\boldmath $\gamma$}}}

\def\bsi{{\text{\boldmath $\sigma$}}}

\def\bpsi{{\text{\boldmath $\psi$}}}

\def\ah{{\hat a}}

\def\Vh{{\widehat V}}

\def\beh{{\widehat \be}}

\def\thh{{\hat \th}}
\def\kah{{\hat \ka}}
\def\lah{{\hat \la}}

\def\sih{{\hat \si}}
\def\tah{{\hat \tau}}

\def\bbeh{{\widehat \bbe}}
\def\bgah{{\widehat \bga}}

\def\bpsih{{\widehat \bpsi}}

\def\wt{{\tilde w}}
\def\yt{{\tilde y}}

\def\ubt{{\tilde \u}}
\def\vbt{{\tilde \v}}

\def\tat{{\tilde \tau}}

\def\lat{{\tilde \la}}

\def\bbet{{\widetilde \bbe}}

\def\bpsit{{\tilde \bpsi}}
\def\co{{\overline c}}

\def\vo{{\overline v}}

\def\zo{{\overline z}}
\def\one{{\bf\text{\boldmath $1$}}}
\def\dd{{\rm d}}
\newcommand{\argmax}{\mathop{\rm argmax}\limits}
\newcommand{\argmin}{\mathop{\rm argmin}\limits}

\def\rN{\mathrm{N}}
\def\IG{\mathrm{IG}}

\def\Gini{\mathrm{GINI}}
\def\WA{\mathrm{WA}}

\begin{document}
\title{Small area estimation of general finite-population parameters based on grouped data}
\date{}

\author{
Yuki Kawakubo\footnote{Graduate School of Social Sciences, Chiba University, 
1-33, Yayoi-cho, Inage-ku, Chiba, 263-8522, Japan,\quad
(E-mail: \texttt{\{kawakubo,gkobayashi\}@chiba-u.jp})} \ and
Genya Kobayashi\footnotemark[1]
}
\maketitle
\begin{abstract}
This paper proposes a new model-based approach to small area estimation of general finite-population parameters based on  grouped data or frequency data, which is often available from sample surveys. 
Grouped data contains information on frequencies of some pre-specified groups in each area, for example the numbers of households in the income classes, and thus provides more detailed insight about small areas than area-level aggregated data.  
A direct application of the widely used small area methods, such as the Fay--Herriot model for area-level data and nested error regression model for unit-level data, is not appropriate since they are not designed for grouped data.
The newly proposed method adopts the multinomial likelihood function for the grouped data. 
In order to connect the group probabilities of the multinomial likelihood and the auxiliary variables within the framework of small area estimation, we introduce the unobserved unit-level quantities of interest which follows the linear mixed model with the random intercepts and dispersions after some transformation. 
Then the probabilities that a unit belongs to the groups can be derived and are used to construct the likelihood function for the grouped data given the random effects. 
The unknown model parameters (hyperparameters) are estimated by a newly developed Monte Carlo EM algorithm using an efficient importance sampling. 
The empirical best predicts (empirical Bayes estimates) of small area parameters can be calculated by a simple Gibbs sampling algorithm. 
The numerical performance of the proposed method is illustrated based on the model-based and design-based simulations. 
In the application to the city level grouped income data of Japan, we complete the patchy maps of the Gini coefficient as well as mean income across the country.

\par\vspace{4mm}
\end{abstract}

\section{Introduction}
\label{sec:int}

Sample surveys are generally designed to estimate finite population parameters, such as total, mean, variance and quantiles.
On the other hand, decision makers of both public and private agencies have become interested in such parameters for smaller subpopulation (small area) as well, created by cross classifying geographical and demographical variables, such as age, sex and race.
However, direct survey estimators of small area parameters, sample mean, sample variance, sample quantiles and others, are often unstable and unreliable because the sample size for each area is  too small mainly due to the budget constraint.
In order to obtain more reliable estimators of small area parameters, the model-based approach which uses mixed effects models is becoming popular.
The empirical best predictor or empirical Bayes estimator derived from mixed effects models, which is often called model based estimator, is more stable than the direct survey estimator because the model-based estimator borrows strength from other areas through the statistical model which connects across the areas with auxiliary variables from other data sources such as large-scale sample surveys and population census.
Alternatively, the hierarchical Bayes approach to the model-based method has been also discussed in the literature.
For the detail about small area estimation (SAE), see \citet{DG12}, \citet{Pfe13}, \citet{RM15} and others.
There are two fundamental models for model-based SAE:  the Fay--Herriot model for area-level aggregated data, which was first proposed to estimate the per capita income for small areas by \citet{FH79}, and the nested error regression model for unit-level data \citep{BHF88}.
While only one population parameter, such as an areal mean, can be estimated at a time by using Fay--Herriot model, general finite population parameters can be estimated by using the nested error regression model and its extensions, proposed by \citet{MR10}, \citet{GMR18}, \citet{DR18}, \citet{SK19} and others provided that a unit-level data is available. 
However, the Fay--Herriot model is more widely used in practice as the accessibility of unit-level data is limited in many cases.

Along with area-level aggregated measures of quantities of interest, as sample mean, sample surveys frequently report grouped data. 
Grouped data contains information on frequency distributions based on  some predefined groups in each area and thus provides more insight about areas than an aggregated areal measure.
The need to model for and to analyze a grouped data arises in many fields of statistical analysis and there exist theoretical developments regarding the grouped data analysis, see \cite{Heitjan89} and references therein. 
Especially in the analysis of income data, the individual households often are grouped into some predefined income classes \citep{Choti08}.
For example, Housing and Land Survey (HLS) conducted by Statistics Bureau of Japan in 2013 reports the numbers of households that fall into the five and nine income classes over 1265 municipalities. 
The grouped data literature, mainly from the view point of the income data analysis, predominantly focused on developing a more flexible underlying parametric or semiparametric form for a single nation, region or period. 
However, when we face the grouped data over multiple local areas as in the HLS data, the existing grouped data methods do not suffice. 
This is because the reported frequency distributions are based on the survey sampling, they are not reliable for areas with small sample sizes and thus call for a correction through an SAE method.  
It must be noted that none of the existing SAE methods can be used to reduce uncertainty in grouped data, because grouped data do not contain unit-level information that is required in the nested error regression model and an appropriate direct estimator that can be used in the Fay--Herriot model is difficult to define for many small area parameters. 
Therefore a new SAE method specifically designed for grouped data is required.

In this paper, we develop a new model-based SAE method which explicitly takes frequency distributions observed in grouped data into account and can estimate general finite population parameters including areal means.
Since the frequency distribution in the grouped data counts the number of units that fall into each group, the multinomial likelihood function is adopted. 
We introduce the latent unit-level variables that represent the unit-level quantities of interest  and that are supported within the range of each group. 
Then in order to connect the frequency distribution to the auxiliary variables within the SAE framework, these latent unit-level variables are assumed to follow a linear mixed model after some transformation.
The linear mixed model adopts the random dispersion as well as random intercept, because the frequency distribution of each area provides the information on the scale of the distribution. 
While \citet{JN12} and \citet{KSGC16} considered the heteroskedasticity in SAE, they did not consider the grouped data setting. 
Given the random effects, the probabilities that a unit belongs to the groups can be derived and are used to construct the multinomial likelihood function  for the grouped data.  
The unknown model parameters (hyperparameters) are estimated by maximizing the marginal likelihood which integrates out the random effects. 
Since the marginal likelihood cannot be evaluated analytically, we develop an EM algorithm \citep{DLR77}, where the E-step is carried out by Monte Carlo integration based on the sampling importance resampling (SIR) using an efficient importance sampling technique.
After obtaining the estimates of hyperparameters, the  empirical Bayes (EB) or equivalently empirical best predicts,  of small area parameters, such as areal means and Gini coefficients, are easily calculated using the output from a simple Gibbs sampler, where the unobserved unit-level quantities are augmented as latent variables to simulate the finite population.

The rest of the paper is organized as follows.
Section~\ref{sec:method} describes the proposed model and methods for hyperparameter estimation and calculation of EB estimates. 
Section~\ref{sec:income} presents the application of the proposed method to Japanese income dataset from HLS. 
The patchy maps of the areal mean income and Gini coefficient are completed using our method. 
In Section \ref{sec:sim}, the performance of the proposed model is examined through the model-based and design-based simulation studies. 
Finally, Section \ref{sec:concl} concludes the paper with some discussion.

\section{Proposed method}
\label{sec:method}

\subsection{Model description}
\label{subsec:model}
In each of $m$ areas, we observe the grouped data that provides the frequency distribution over the mutually exclusive $G$ groups divided by the known thresholds $0 = c_0 < c_1 < \dots < c_{G-1} < c_G = +\infty$. 
Let us denote the observed frequencies and sample size in the $i$th area by $\y_i = (y_{i1},\dots,y_{iG})^\top$ for $i=1,\dots,m$ and  $n_i=\sum_{g=1}^G y_{ig}$, respectively, and thus $y_{ig}$ counts the number of units that fall into the $g$th group in the $i$th area. 
Therefore, it can be regarded that $\y_i$ follows the multinomial distribution. 
In order to model the group probabilities of the multinomial distribution that links the grouped data with the auxiliary variables and then to facilitate the small area parameter estimation (see Section~\ref{subsec:Gibbs}), we introduce the positive latent variable $z_{ij}>0$ for the $j$th unit in the $i$th area  ($i=1,\dots,m; \ j=1,\dots,N_i$) that constitutes the population of the $i$th area and from which the units are sampled to construct the grouped data.  
We also let $\z_i = (z_{i1},\dots,z_{iN_i})^\top$. 
Note that $N_i$ is not the sample size but the population size and thus a finite population setting is considered. 
Without loss of generality, it is assumed that the first $n_i$ values of $z_{ij}$'s are sampled. 
Then $y_{ig}$ can be  expressed as 
\begin{equation}
\label{eqn:model_y}
y_{ig} = \sum_{j=1}^{n_i} I( c_{g-1} \leq z_{ij} < c_g ), \quad (g=1,\dots,G), \\
\end{equation}
where $I(\cdot)$ is the indicator function. 
We take into account the variability of the frequency distribution by incorporating the sample size into our model.

In order to devise small area estimation for the grouped data, we assume that the latent $z_{ij}$ after some transformation follows the linear mixed model:
\begin{equation}
\label{eqn:lmm}
\begin{split}
&h_\ka(z_{ij}) = \x_i^\top\bbe + b_i + \ep_{ij}, \quad b_i \sim \rN(0,\tau^2), \\
&\ep_{ij} \mid \si_i^2 \sim \rN(0,\si_i^2), \quad \si_i^2 \sim \IG \left( {\la \over 2} + 1, {\la\varphi_i \over 2} \right), \quad \varphi_i = \exp(\x_i^\top\bga),
\end{split}
\end{equation}
or equivalently the following Bayesian model:
\begin{equation}
\label{eqn:BM}
\begin{split}
h_\ka(z_{ij}) \mid \mu_i, \si_i^2 &\sim \rN(\mu_i,\si_i^2) \\
\mu_i &\sim \rN(\x_i^\top\bbe, \tau^2) \\
\si_i^2 &\sim \IG\left( { \la \over 2 }+ 1, {\la\varphi_i \over 2} \right), \quad \varphi_i = \exp(\x_i^\top\bga),
\end{split}
\end{equation}
where $h_\ka(\cdot)$ is an arbitrary parametric transformation with the parameter $\kappa$,
$\x_i$ is the area specific $p$-dimensional auxiliary variable vector, $\bbe$ is the unknown parameter vector of regression coefficients, $b_i$ is the random area effect with the unknown variance parameter $\tau^2$ and $\ep_{ij}$ is the error term with the area specific random variance $\sigma^2_i$. 
It is further assumed that $b_i$'s and $\si_i^2$'s are mutually independent or equivalently $\mu_i$'s and $\si_i^2$'s are mutually independent and that $z_{ij}$'s are conditionally independent given $\b = (b_1,\dots,b_m)^\top$ and $\bsi = (\si_1^2,\dots,\si_m^2)^\top$.
The mean of $\sigma_i^2$ is $\varphi_i$ which is further modeled as $\varphi_i=\exp(\x_i^\top\bga)$ using the auxiliary variables.
While the model looks like a version of unit-level nested error regression model proposed in the small area estimation literature \citep{BHF88},
there is a crucial difference that in the present setting we do not observe the unit-level $\z_{i}$'s but $\y_i$'s only.  
Also, the auxiliary variables $\x_i$ are available only at the area-level.

Based on the statistical model \eqref{eqn:lmm} or \eqref{eqn:BM}, the conditional probability that $z_{ij}$ falls in the $g$th group given $b_i$ (or $\mu_i$) and $\si_i^2$ is given by
\begin{equation}
\label{eqn:prob_g}
\Pr(c_{g-1} \leq z_{ij} < c_g \mid b_i,\sigma^2_i) =
\Phi\left\{\frac{h_\ka(c_g) - \mu_i}{\sigma_i}\right\}-\Phi\left\{\frac{h_\ka(c_{g-1})-\mu_i}{\sigma_i}\right\},
\end{equation}
where $\mu_i = \x_i^\top \bbe + b_i$ and $\Phi(\cdot)$ denotes the cumulative distribution function of the standard normal distribution.


Note that we model the unit-level variable $z_{ij}$, not the area-level variable like the Fay--Herriot model.
However, the auxiliary variables are available only on the area-level. 
Hence, if the log transformation is used,  the superpopulation of $z_{ij}$ is the log-normal distribution with the same mean and variance within the same small area $i$, which is too restrictive.
In this paper, a more flexible parametric transformation $h_\ka(\cdot)$ is adopted to relax the restriction. 
Specifically, we use the Box--Cox transformation given by
\begin{equation*}
h_\ka(z)=\left\{
\begin{split}
\frac{z^\kappa-1}{\kappa},\quad \kappa\neq 0,\\
\log(z),\quad \kappa=0,
\end{split}
\right. \quad z>0,
\end{equation*}
and  $-1/\kappa < h_\kappa(z) < +\infty$ if $\kappa > 0$ and $-\infty < h_\kappa(z) < -1/\kappa$ if $\kappa < 0$.


Our goal is to estimate (predict) some characteristics of each area, such as the areal mean $\zo_i = N_i^{-1}\sum_{j=1}^{N_i}z_{ij}$ and Gini coefficient defined as
\begin{equation}
\label{eqn:Gini}
\mathrm{GINI}(\z_i) = {1 \over N_i} \left\{ N_i + 1 - {2\sum_{j=1}^{N_i}(N_i + 1 - j)z_{i(j)} \over N_i\zo_i} \right\},
\end{equation}
where $\{ z_{i(1)},\dots, z_{i(N_i)} \}$ are sorted values of $\{ z_{i1},\dots, z_{i,N_i} \}$ in non-decreasing order.
To this end, we develop the empirical Bayes (EB) estimators of $\zo_i$ and $\Gini(\z_i)$.

\subsection{Hyperparameter estimation}
\label{subsec:EM}
The unknown model parameter vector is denoted by $\bpsi = (\bbe^\top, \ta^2, \la, \ka, \bga^\top)^\top$.
If our model is seen as a Bayesian model  \eqref{eqn:BM}, $\bpsi$ is referred to as hyperparameters.
Hereafter, $\bpsi$ is referred to as the hyperparmeters for the sake of clarity of terminology. 

The hyperparameter $\bpsi$ is estimated by maximizing the marginal likelihood:
\begin{equation}
\label{eqn:ML}
L( \bpsi; \y ) = \prod_{i=1}^m \int f( \y_i \mid \u_i) \pi(\u_i) \dd \u_i,
\end{equation}
where $\pi(\u_i)$ is the pdf of $\u_i = (b_i, \si_i^2)^\top \sim \rN(0,\tau^2) \times \IG(\lambda/2+1, \lambda\varphi_i/2)$, and $f(\y_i \mid \u_i)$ is the conditional probability mass function (pmf) of $\y_i$ given $\u_i$, which is given by the pmf of the multinomial distribution with $n_i$ trials and the probabilities given by \eqref{eqn:prob_g}:
\begin{equation}
\label{eqn:pmf_yi}
f( \y_i \mid \u_i ) = { n_i! \over y_{i1}!y_{i2}!\cdots y_{iG}! } \times \prod_{g=1}^G \left[ \Phi\left\{ \frac{ h_\ka(c_g) - \mu_i }{ \si_i} \right\} - \Phi\left\{ \frac{ h_\ka(c_{g-1}) - \mu_i }{ \si} \right\} \right]^{y_{ig}},
\end{equation}
for $i = 1,\dots,m$.
It is difficult to evaluate the marginal likelihood \eqref{eqn:ML} analytically because of the integration with respect to $\u_i$.
Thus we introduce the EM algorithm \citep{DLR77} where  the vector of random effects $\u = (\u_1^\top,\dots,\u_m^\top)^\top$ is regarded as the missing variable.
The complete log-likelihood is given by
\begin{equation*}
\log \{ L^c( \bpsi ; \y, \u ) \} = \sum_{i=1}^m \left[ \log \{ f( \y_i \mid \u_i ) \} + \log \{ \pi(\u_i) \} \right].
\end{equation*}
In the $k$th iteration of the algorithm,  the E-step calculates
\begin{equation*}
Q(\bpsi \mid \bpsi^{(k-1)}) = E[ \log\{ L^c( \bpsi; \y, \u ) \} \mid \y, \bpsi^{(k-1)} ],
\end{equation*}
where the expectation is taken with respect to the conditional distribution of $\u$ given $\y$ with the parameter value $\bpsi^{(k-1)}$ from the $(k-1)$th iteration.
The M-step  maximizes $Q(\bpsi \mid \bpsi^{(k-1)})$ with respect to $\bpsi$. 
The  maximizer, denoted by $\bpsi^{(k)} = ( ( \bbe^{(k)} )^\top, \tau^{2(k)}, \la^{(k)}, \kappa^{(k)},(\bga^{(k)})^\top )^\top$, is obtained as
\begin{align*}
\tau^{2(k)} =& \ {1 \over m}E[ \b^\top\b \mid \y, \bpsi^{(k-1)} ], \\
( ( \bbe^{(k)} )^\top, \kappa^{(k)} )^\top =& \ \argmax_{ ( \bbe^\top, \ka )^\top } E\left[ \sum_{i=1}^m \log\{ f( \y_i \mid \u_i ) \} \bigm| \y, \bpsi^{(k-1)} \right], \\
( (\bga^{(k)})^\top, \la^{(k)} )^\top =& \ \argmax_{ ( \bga^\top, \la )^\top } E\left[ \sum_{i=1}^m \log\{ \pi(\si_i^2) \} \bigm| \y, \bpsi^{(k-1)} \right].
\end{align*}

Since it is difficult to evaluate the conditional expectation analytically in the E-step,  we use the Monte Carlo integration based on the sampling importance resampling (SIR).
Note that the conditional pdf of $\u$ given $\y$ is the product of the conditional pdfs of $\u_i$ given $\y_i$: 
\begin{equation*}
\pi( \u \mid \y ) = \prod_{i=1}^m \pi( \u_i \mid \y_i )\propto\prod_{i=1}^m f(\y_i \mid \u_i) \pi(\u_i),
\end{equation*}
where $\pi( \u \mid \y )$ is the conditional pdf of $\u$ given $\y$ and $\pi( \u_i \mid \y_i )$ is the conditional pdf of $\u_i$ given $\y_i$.
Therefore, we apply the following SIR method independently for $i=1,\dots,m$. 
Let $q(\u_i \mid \a_i)$ denote the proposal density for $\u_i$ where $\a_i\in\mathbb{R}^q$ is the parameter vector of the proposal distribution. 
In the SIR method, first a set of random numbers $\{ \ubt_i^{(1)},\dots,\ubt_i^{(S_1)} \}$ from $q(\u_i \mid \a_i)$ is generated. 
Then for each $\ubt_i^{(s)}$, the weight 
\begin{equation*}
\wt_{is}=\frac{f(\y_i \mid \ubt_i^{(s)}) \pi(\ubt_i^{(s)}) }{ q( \ubt_i^{(s)} \mid \a_i) },\quad s=1,\dots,S_1,
\end{equation*}
is calculated. 
Finally, a set of samples of size $S_2$, $\{ \u_i^{(1)},\dots,\u_i^{(S_2)} \}$, is drawn with replacement from $\{ \ubt_i^{(1)},\dots,\ubt_i^{(S_1)} \}$ based on the probability 
\begin{equation*}
\Pr( \u_i^{(r)} = \ubt_i^{(s)} ) = \frac{\wt_{is}}{\sum_{s'=1}^{S_1} \wt_{is'}},\quad s=1,\dots,S_1,\quad r=1,\dots,S_2. 
\end{equation*}
For large $S_1/S_2$, $\{ \u_i^{(1)},\dots,\u_i^{(S_2)} \}$ is approximately a set of independent random samples from $\pi( \u_i \mid \y_i )$. 
The expectations in the M-step are replaced with the Monte-Carlo estimates based on the SIR samples.

The performance of the SIR depends on the choice of the proposal distribution. 
It is ideal to employ a proposal distribution that well approximates the target distribution and we aim to achieve this by updating the value of $\a_i$ through an iterative procedure proposed by \citet{RZ07}. 
Their efficient importance sampling (EIS) method determines the value $\hat{\a}_i$ such that it minimizes the Monte Carlo sampling variance of the importance weights with respect to the proposal distribution. 
In the current context, as shown by \citet{RZ07},  $\hat{\a}_i$ is determined through the following minimization problem
\begin{equation}\label{eqn:eis_q}
(\hat{c}_i, \hat{\a}_i^\top)^\top = \argmin_{( c_i,\a_i^\top)^\top}\int \left\{\log f(\y_i \mid \u_i) + \log\pi(\u_i) - c_i - \log g(\u_i \mid \a_i)\right\}^2 w_i(\u_i \mid \a_i) q(\u_i \mid \a_i) \dd\u_i,
\end{equation}
where $g(\u_i \mid \a_i)$ is the kernel of the proposal density $q( \u_i \mid \a_i)$ such that $q(\u_i \mid \a_i)=g(\u_i \mid \a_i)/\int g(\u_i \mid \a_i) \dd \u_i$, $c_i$ is a scalar that adjusts for the normalizing constants and $w_i(\u_i \mid \a_i)=f(\y_i \mid \u_i)\pi(\u_i)/q(\u_i \mid \a_i)$.  
The EIS method replaces \eqref{eqn:eis_q} with a Monte Carlo approximation and proceeds by iteratively solving
\begin{equation}\label{eqn:eis_qmc}
(\hat{c}_i^{(t)},\hat{\a}_i^{(t)\top})^\top = \argmin_{( c_i,\a_i^\top)^\top}\frac{1}{S_0}\sum_{s=1}^{S_0} \left\{\log f(\y_i \mid \check{\u}_i^{(s)})+\log\pi(\check{\u}_i^{(s)}) - c_i - \log g(\check{\u}_i^{(s)} \mid \a_i)\right\}^2 w_i(\check{\u}_i^{(s)} \mid \a_i^{(t-1)}),
\end{equation}
where $(\hat{\a}_i^{(t)\top}, \hat{c}_i^{(t)})^\top$ denotes the value of $(\hat{\a}_i^\top, \hat{c}_i)^\top$ at the $t$th iteration of the EIS minimization and $\{ \check{\u}_i^{(1)},\dots,\check{\u}_i^{(S_0)} \}$ is the set of samples generated from $q(\u_i \mid \hat{\a}_i^{(t-1)})$ for $\check{\u}_i^{(s)} = ( \check{b}_i^{(s)}, \check{\si}_i^{2(s)} )^\top$. 
\citet{RZ07} noted that $S_0$ does not have to be very large. 
In this paper, we employ $\rN(\th_{i1}(\a_i),\th_{i2}(\a_i)) \times \IG( \th_{i3}(\a_i), \th_{i4}(\a_i) )$ for $q(\u_i \mid \a_i)$ where $\a_i = (a_{i1}, a_{i2}, a_{i3}, a_{i4})^\top$ is the vector of natural parameters.
Because the proposal distribution belongs to the exponential family where
$$
\log g(\u_i^{(s)} \mid \a_i) = a_{i1}b_i + a_{i2}b_i^2 + a_{i3} \log(\si_i^2) + a_{i4}{1 \over \si_i^2},
$$
for $a_{i1} = \th_{i1} / \th_{i2}$, $a_{i2} = -1 / (2\th_{i2})$, $a_{i3} = -(\th_{i3} + 1)$ and $a_{i4} = -\th_{i4}$, the solution for the EIS minimization \eqref{eqn:eis_qmc} is given by the following generalized least squares (GLS) estimator
\begin{equation}\label{eqn:eis_gls}
(\hat{c}_i^{(t)}, \hat{\a}_{i}^{(t)\top})^\top=(\Z_i^\top\D_i\Z_i)^{-1}\Z_i^\top\D_i\f_i
\end{equation}
where 
$\Z_i=(\one_{S_0},\check{\b}_i,\check{\b}_i^{2}, \mathbf{log}\check{\bsi}_i^2, \check{\bsi}_i^{-2} )$,  $\check{\b}_i$, $\check{\b}_i^2$, $\mathbf{log}\check{\bsi}_i^2$, $\check{\bsi}_i^{-2}$ and $\f_i$ are $S_0\times 1$ vectors with the $s$th elements given by $\check{b}_i^{(s)}$, $(\check{b}_i^{(s)})^2$, $\log(\check{\si}_i^{2(s)})$, $1 / \check{\si}_i^{2(s)}$ and $\log f(\y_i \mid \check{\u}_i^{(s)})+\log\pi(\check{\u}_i^{(s)})$, respectively,
and $\D_i$ is the $S_0$ dimensional diagonal matrix with $w_i(\check{\u}_i^{(s)} \mid \hat{\a}_i^{(t-1)})$ on the $s$th diagonal position.  
In this paper, the EIS iteration is terminated when the relative change in $(\th_{i1}(\a_i^{(t)}),\th_{i2}(\a_i^{(t)}), \th_{i3}(\a_i^{(t)}), \th_{i4}(\a_i^{(t)}) )^\top$ is below $10^{-3}$. 
After the termination of the EIS iterations, the optimal parameters for the proposal distribution are obtained through $\thh_{i1}=-\hat{a}_{1i}/(2\hat{a}_{2i})$, $\thh_{i2}=-1/(2\hat{a}_{2i})$, $\thh_{i3} = -\ah_{3i} - 1$ and $\thh_{i4} = -\ah_{4i}$. 
See \citet{RZ07} for more detailed implementation of the EIS method. 

The initial values for the MCEM algorithm are determined as follows.
Let us define $V_i = n_i^{-1}\sum_{g=1}^G\log(\co_g) \times y_{ig}$ where $\co_g = ( c_{g-1} + c_g ) / 2$ for $g=1,\dots,G-1$ and $\co_G = c_{G-1} + ( c_{G-1} - c_{G-2} ) / 2$,  $\V = (V_1,\dots,V_m)^\top$ and $\X = (\x_1,\dots,\x_m)^\top$. 
Then, the initial value of $\bbe$ and $\ta^2$ are determined as
$$
\bbe^{(0)} = (\X^\top\X)^{-1}\X^\top\V, \quad \tau^{2(0)} = m^{-1}\Vert \V - \X\bbe^{(0)} \Vert^2.
$$
The initial values of $\la$, $\kappa$ and $\bga$ are determined by using the estimates based on the local model which modifies the model \eqref{eqn:lmm} as follows:
\begin{equation}
\label{eqn:local}
h_{\kappa_i}(z_{ij}) = \beta_i + \ep_{ij}, \quad \ep_{ij} \sim \rN(0,\si_i^2),
\end{equation}
where $\be_i,$, $\kappa_i$ and $\si_i^2$ are the unknown parameters. 
Let $\beh_i$, $ \kah_i$ and $\sih_i^2$ denote the maximum likelihood estimates which independently maximizes the likelihood function for $i=1,\dots,m$:
$$
(\beh_i, \kah_i, \si_i^2)^\top = \argmax_{(\be_i,\ka_i,\si_i^2)^\top}{ n_i! \over y_{i1}!y_{i2}!\cdots y_{iG}! } \times \prod_{g=1}^G \left[ \Phi\left\{ \frac{ h_\ka(c_g) - \be_i }{ \si_i} \right\} - \Phi\left\{ \frac{ h_\ka(c_{g-1}) - \be_i }{ \si} \right\} \right]^{y_{ig}}. 
$$
Then, the initial value of $\la$ and $\kappa$ are determined as
$$
\la^{(0)} = 2 \times \{ (\overline{\sih^2})^2 / \Vh(\sih^2) +1 \}, \quad \kappa^{(0)} = \overline{\kah},
$$
where $\overline{\sih^2}$ and $\Vh(\sih^2)$ are sample mean and variance of $\sih_i^2$'s over the areas and $\overline{\kah}$ is the sample mean of $\kah_i$'s.
Furthermore, the initial value of $\bga$ is
$$
\bga^{(0)} = (\X^\top\X)^{-1}\X^\top \bsi,
$$
where $\bsi = (\sih^2_1,\dots,\sih^2_m)^\top$.
This method generally provides reasonable initial values for the MCEM algorithm leading to a fast convergence. 
Although other initial values are also tried, the similar results are obtained with longer computing times.

To monitor the convergence  of the MCEM algorithm, the criterion considered by \cite{SC02} is used. 
In order to prevent premature termination of the algorithm due to the difference in the scale of the parameter values, the quantities $e_{k,(\bbe)}$, $e_{k,(\tau^2)}$, $e_{k,(\kappa)}$, $e_{k,(\lambda)}$ and $e_{k,(\bga)}$ is evaluated respectively for $\bbe$, $\tau^2$, $\kappa$, $\lambda$ and $\bga$. 
In the case of $\bbe$, for example, 
\begin{equation}\label{eqn:em_conv}
e_{k,(\bbe)} = { \| \bbet_1^{(k)} - \bbet_2^{(k)} \| \over \| \bbet_2^{(k)} \| + \de },
\end{equation}
where $\bbet_1^{(k)} = H^{-1} \sum_{h=0}^{H-1} \bbe^{(k-h)}$, $\bbet_2^{(k)} = H^{-1} \sum_{h=0}^{H-1} \bbet^{(k-h-d)}$, and $\de$, $H$, and $d$ are specified by the user. 
Then the EM algorithm is terminated in the $k$th iteration if 
$$
\max\{ e_{k,(\bbe)}, e_{k,(\tau^2)}, e_{k,(\kappa)}, e_{k,(\lambda)}, e_{k,(\bga)}\} < \epsilon,
$$
for some small value $\epsilon>0$, and use $\bpsit_1^{(k)} = ( \bbet^{(k)\top}, \tat^{2(k)}, \lat^{(k)}, \tilde{\ka}^{(k)}, \tilde{\bga}^{(k)\top} )^\top$ as the estimate of $\bpsi$, which is denoted by $\bpsih = (\bbeh^\top, \tah^2, \lah, \kah, \bgah^\top)^\top$ hereafter.

\subsection{Calculation of empirical Bayes estimates}
\label{subsec:Gibbs}

Here we propose the method to calculate EB estimates of some function of $\z_i$, which is denoted as $\zeta_i(\z_i)$ in general.
The examples of $\zeta_i(\z_i)$ include the areal mean $\zo_i$ and Gini coefficients $\Gini(\z_i)$  in \eqref{eqn:Gini}.
Under the quadratic loss, the Bayes estimator of $\zeta_i(\z_i)$ is its conditional expectation given the data,  $E[ \zeta_i(\z_i) \mid \y ]$.
Because of the independence over the areas, $E[ \zeta_i(\z_i) \mid \y ]$ is reduced to $E[ \zeta_i(\z_i) \mid \y_i ]$, which is denoted by
$$
\xi_i(\bpsi;\y_i) = E[ \zeta_i(\z_i) \mid \y_i ].
$$
Because $\xi_i(\bpsi;\y_i)$ is a function of the unknown parameter $\bpsi$, we obtain the empirical Bayes (EB) estimator $\xi_i(\bpsih;\y_i)$ by substituting $\bpsih$ for $\bpsi$ in the Bayes estimator.
However, since it is impossible to evaluate the conditional expectation of $\zeta_i(\z_i)$ analytically, we calculate the EB estimates from the output of the following Gibbs sampler.


Let the random vector $\vbt_i = (v_{i1},\dots,v_{in_i})^\top$ denote the sorted values of $\{ h_\kah(z_{i1}),\dots,h_\kah(z_{in_i}) \}$ in increasing order with size $y_{i1},\dots,y_{iG}$ and then the following relationship holds:
\begin{equation*}
v_{ij} \leq v_{ik}, \quad {\rm for \ all} \ j,k \ {\rm such \ that} \ j \leq \yt_{ig} < k, \ {\rm for \ all} \ g=1,\dots,G,
\end{equation*}
where $\yt_{ig} = \sum_{g'=1}^g y_{ig'}$ for $g=1,\dots,G$ and $n_i = \yt_{iG}$.
For out-of-sample units, let $\check{\v}_i = (v_{i,n_i+1},\dots,v_{iN_i})^\top = ( h_\kah(z_{i,n_i+1}),\dots,h_\kah( z_{iN_i} ) )^\top$.
Let $\v_i = ( \vbt_i^\top, \check{\v}_i^\top )^\top = (v_{i1},\dots,v_{iN_i})^\top$.
To evaluate the conditional expectation of $\v_i$ given $\y_i$, the sample from the joint conditional distribution of $\{ \vbt_i, \check{\v}_i, \mu_i, \si_i^2 \}$ given $\y_i$ is obtained by using the Gibbs sampling algorithm with the following full conditional distributions: 
\begin{equation}
\label{eqn:full}
\begin{split}
\mu_i \mid \vbt_i, \check{\v}_i, \si_i^2, \y_i &\sim \rN \left( { \si_i^2\x_i^\top\bbeh + N_i\tah^2 \vo_i \over \si_i^2 + N_i\tah^2 }, {\tah^2\si_i^2 \over \si_i^2 + N_i\tah^2} \right), \\
v_{ij} \mid \mu_i, \check{\v}_i, \si_i^2, \y_i & \overset{ \text\small\rm{indep} }{\sim}
\begin{cases}
{\rm TN}_{[ h_\kah(c_0), h_\kah(c_1) )} ( \mu_i, \si_i^2 ),  & j=1,\dots,\tilde{y}_{i1}, \\
{\rm TN}_{[ h_\kah(c_1), h_\kah(c_2) )} ( \mu_i, \si_i^2 ),   & j=\tilde{y}_{i1}+1,\dots,\tilde{y}_{i2} \\
\vdots \\
{\rm TN}_{[ h_\kah(c_{G-1}), h_\kah(c_G) )} ( \mu_i, \si_i^2 ), & j=\tilde{y}_{i,G-1}+1,\dots,n_i,
\end{cases}\\
\check{\v}_i \mid \mu_i, \vbt_i, \si_i^2, \y_i &\sim \mathrm{N}_{N_i-n_i}( \mu_i\one_{N_i - n_i}, \si_i^2\one_{N_i - n_i} ), \\
\si_i^2 \mid \mu_i, \vbt_i, \check{\v}_i, \y_i &\sim \IG\bigg( {N_i+\lah \over 2} + 1, \ {1\over 2} \Big\{ \lah\hat{\varphi}_i + \sum_{j=1}^{N_i} (v_{ij} - \mu_i)^2 \Big\} \bigg),
\end{split}
\end{equation}
where $\vo_i = N_i^{-1}\sum_{j=1}^{N_i}v_{ij}$ and $\mathrm{TN}_{[a,b)}(\mu,\si^2)$ denotes the truncated normal distribution with the mean $\mu$ and variance $\si^2$ truncated to the interval $[a,b)$.
The derivation of the full conditional distributions is given in Appendix~\ref{sec:app1}.

Let $\v_i^{(s)} = (v_{i1}^{(s)},\dots,v_{iN_i}^{(s)})^\top$ be the $s$th output of $\v_i$ from the Gibbs sampler $(s=1,\dots,S_3)$.
Then the EB estimates $\xi_i(\bpsih;\y_i)$ can be calculated as
$$
\widehat{\xi_i(\bpsih;\y_i)} = {1 \over S_3}\sum_{s=1}^{S_3} \zeta_i( h_\kah^{-1}(\v_i^{(s)}) ),
$$
where $h_\kah^{-1}(\cdot)$ is the inverse Box--Cox transformation with parameter value $\kah$.

If the auxiliary variables $\x_i$'s are available for out-of-sample areas,  $\zeta_i(\z_i)$ can be also predicted for an out-of-sample area $i=m+1$  by $\xi_{m+1}(\bpsih)$ where $\xi_{m+1}(\bpsi) = E[ \zeta_{m+1}(\z_{m+1}) ]$, since $\y$ and $z_{m+1}$ are mutually independent. 
This expectation can be calculated by the Monte Carlo integration that generates random numbers from the model \eqref{eqn:lmm} with the hyperparameters are fixed to their estimates.


\section{Application to grouped income data of Japan}
\label{sec:income}
\begin{figure}[H]
\center
\begin{tabular}{ccc}
\includegraphics[scale=0.14]{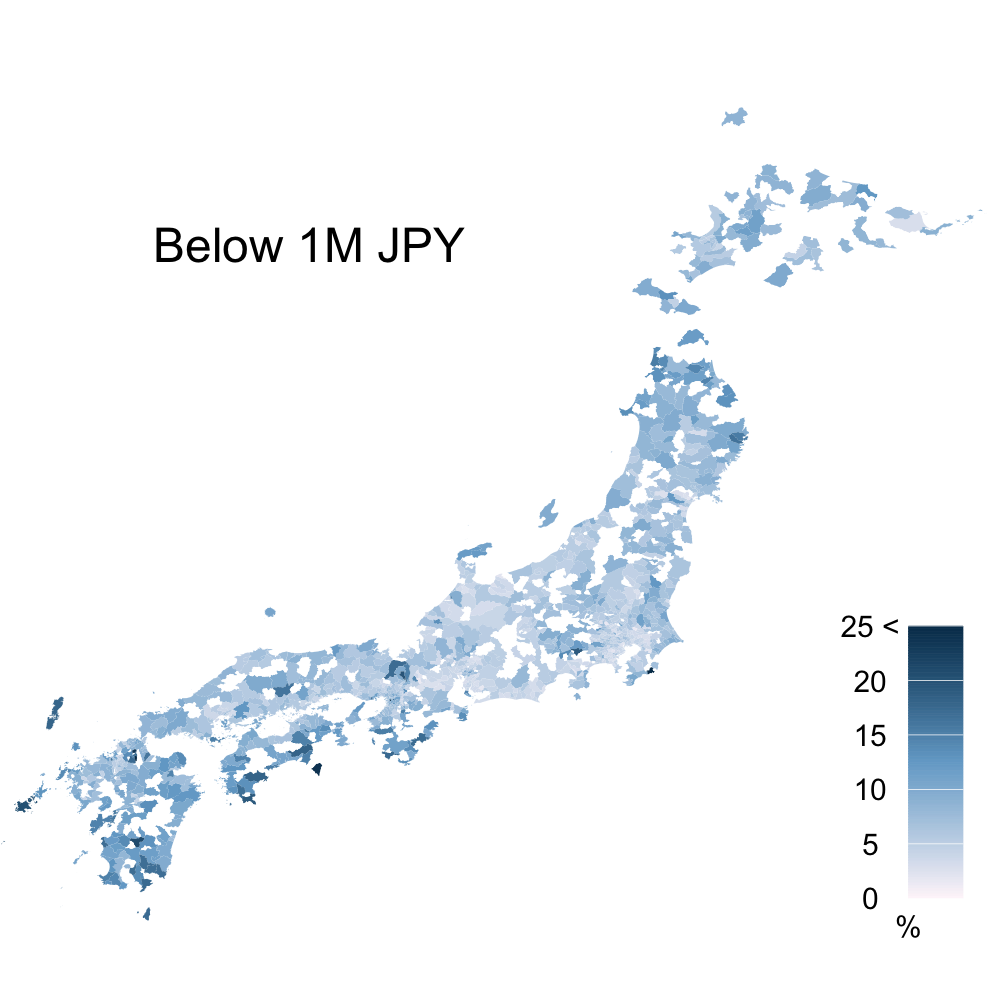} &
\includegraphics[scale=0.14]{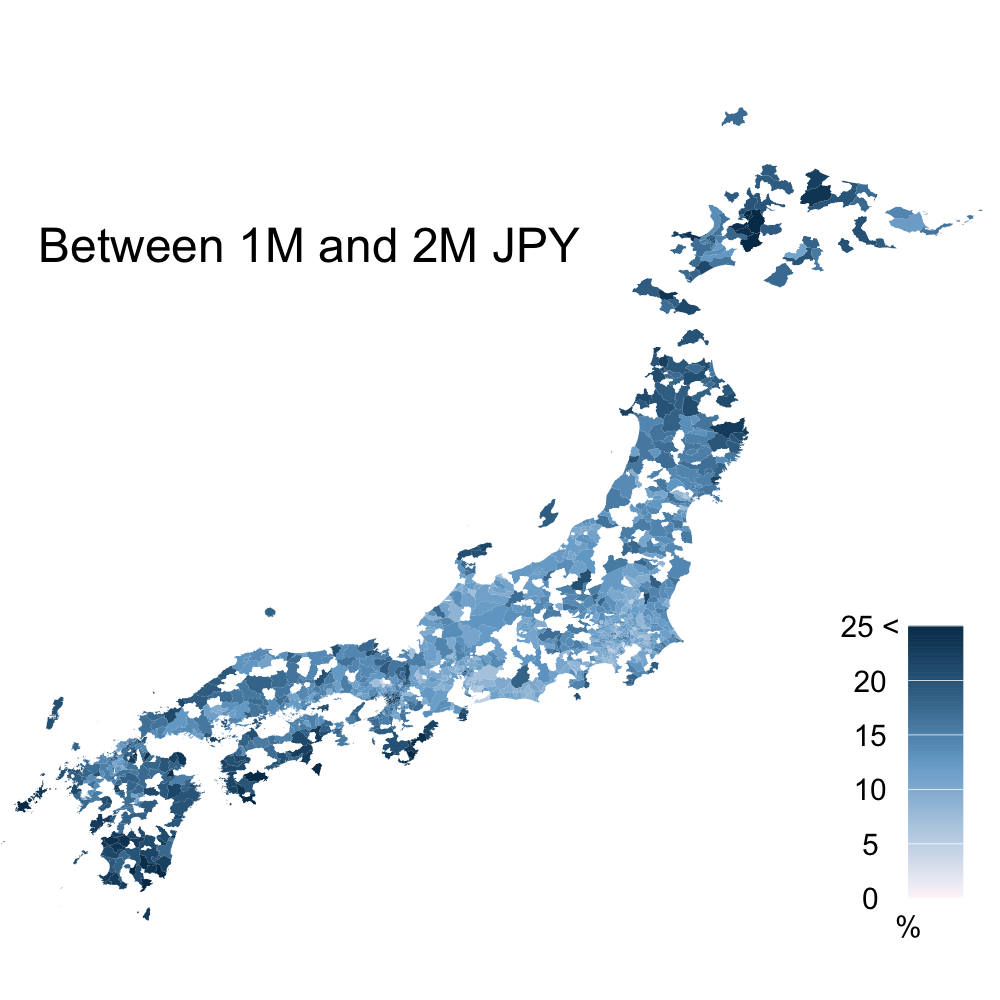} & 
\includegraphics[scale=0.14]{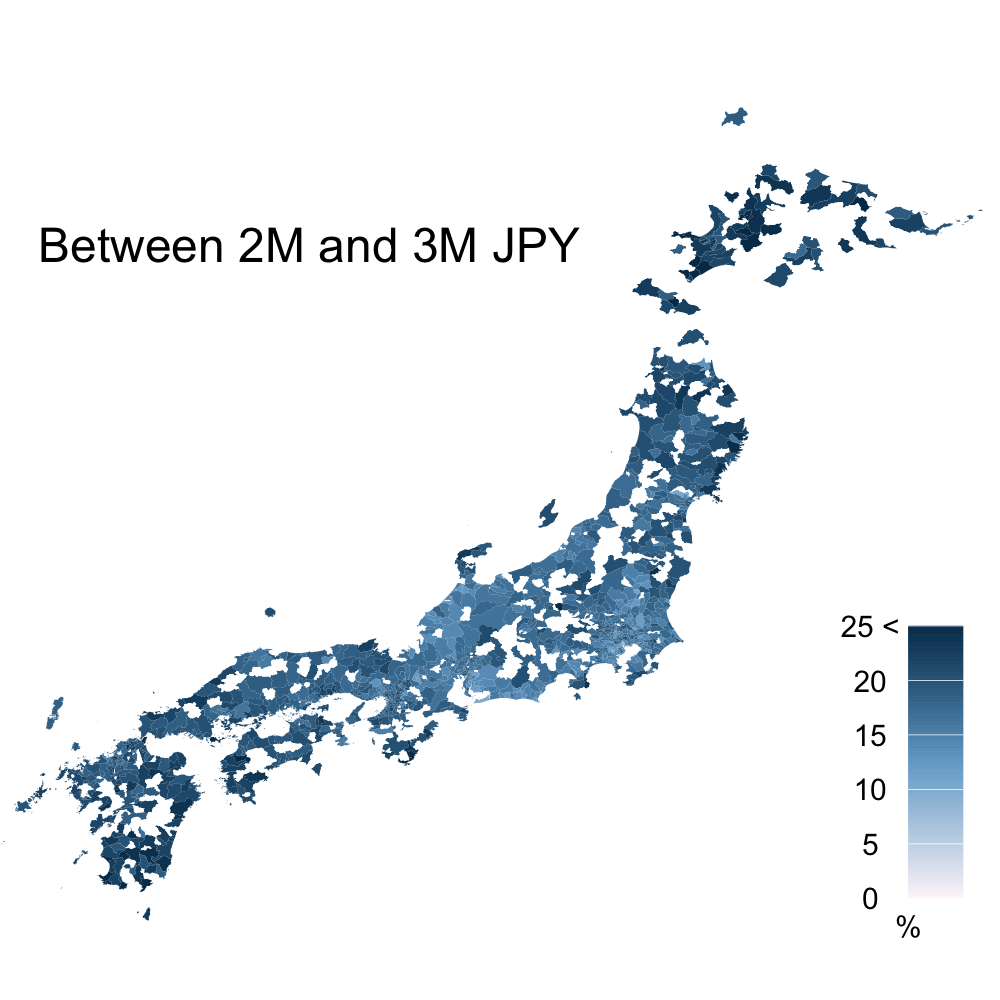} \\
\includegraphics[scale=0.14]{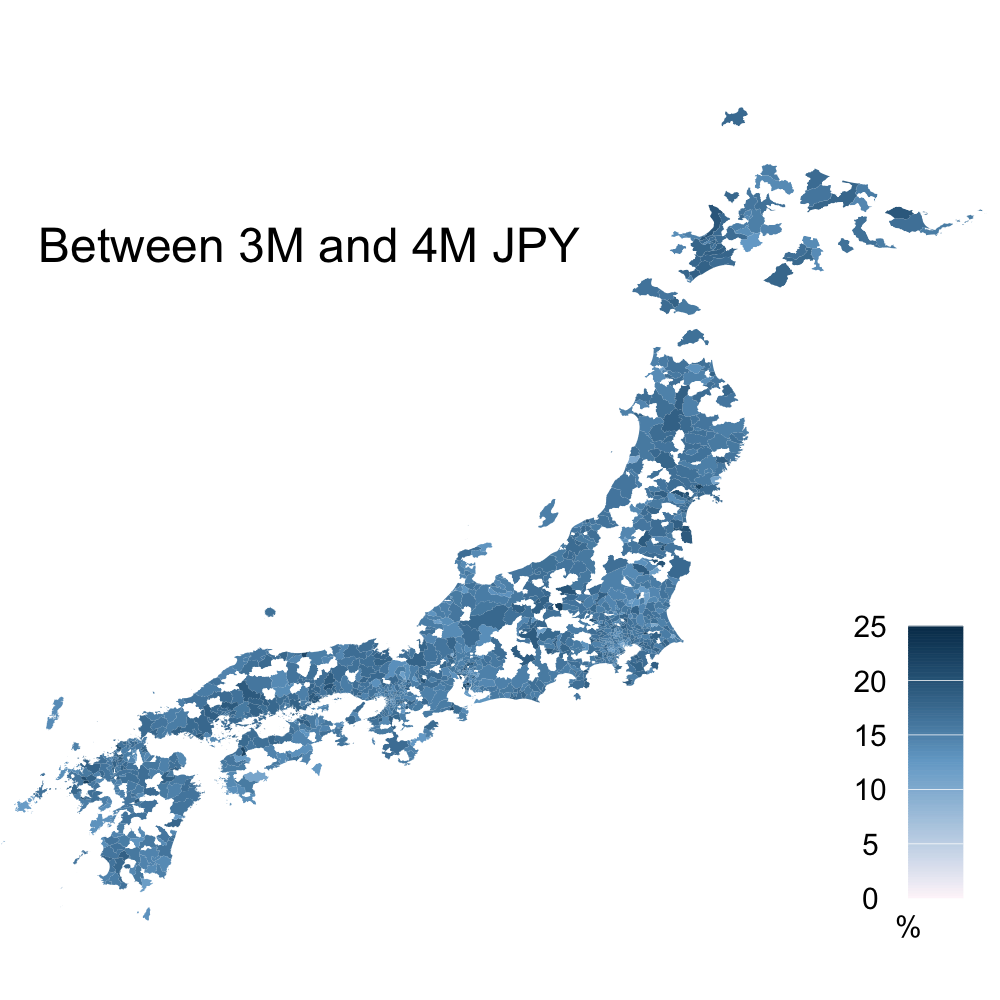} &
\includegraphics[scale=0.14]{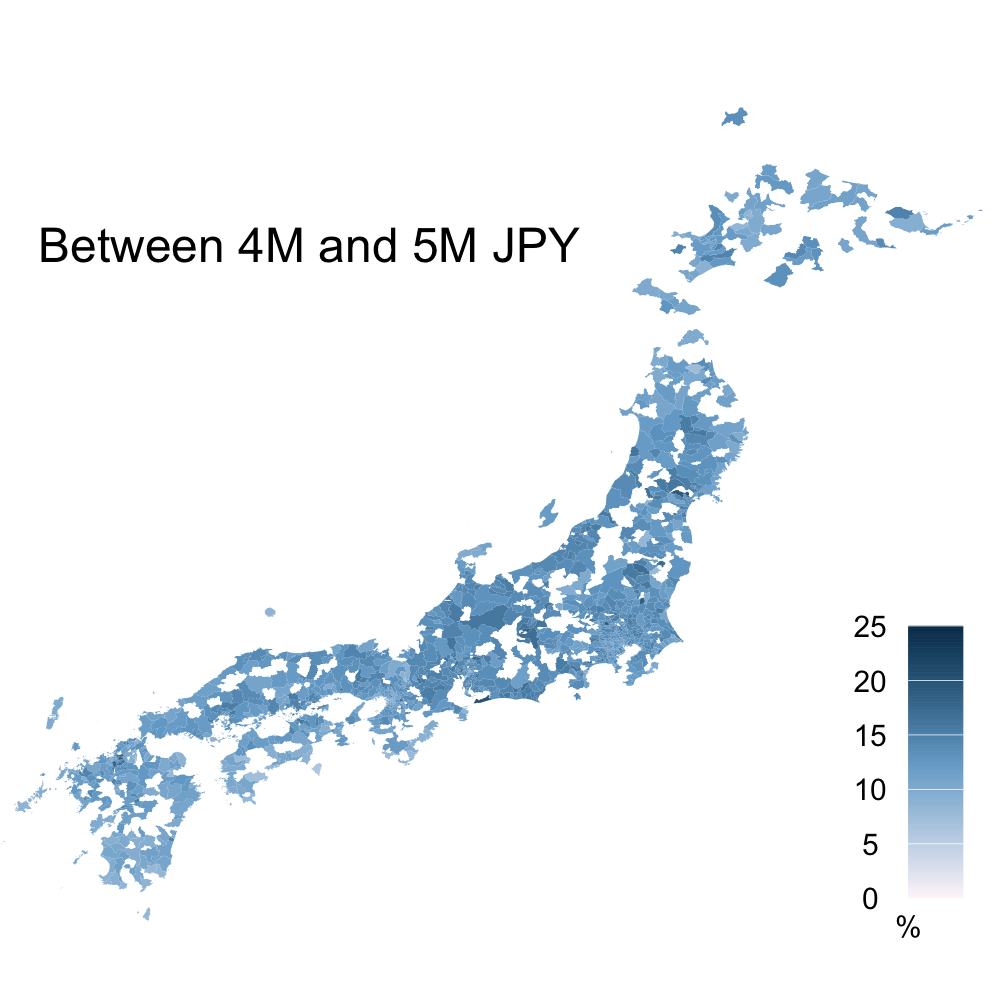} &
\includegraphics[scale=0.14]{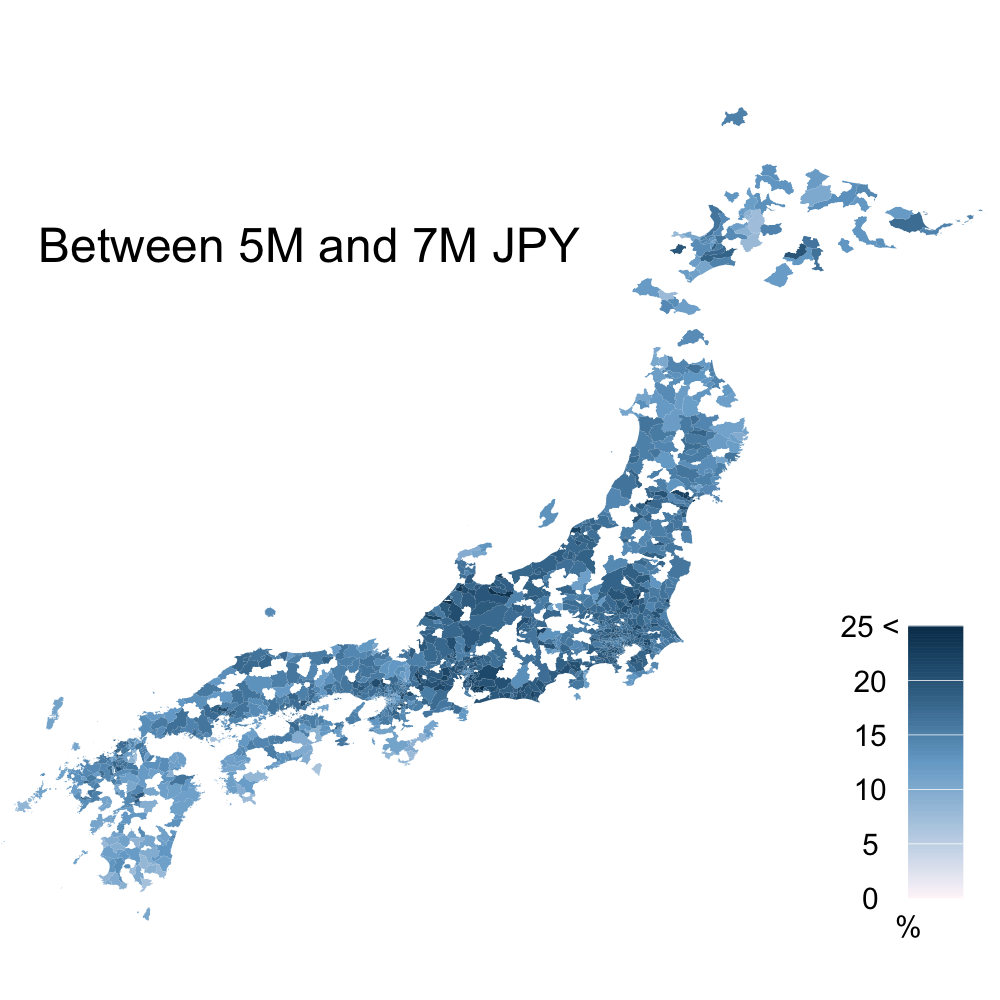} \\
\includegraphics[scale=0.14]{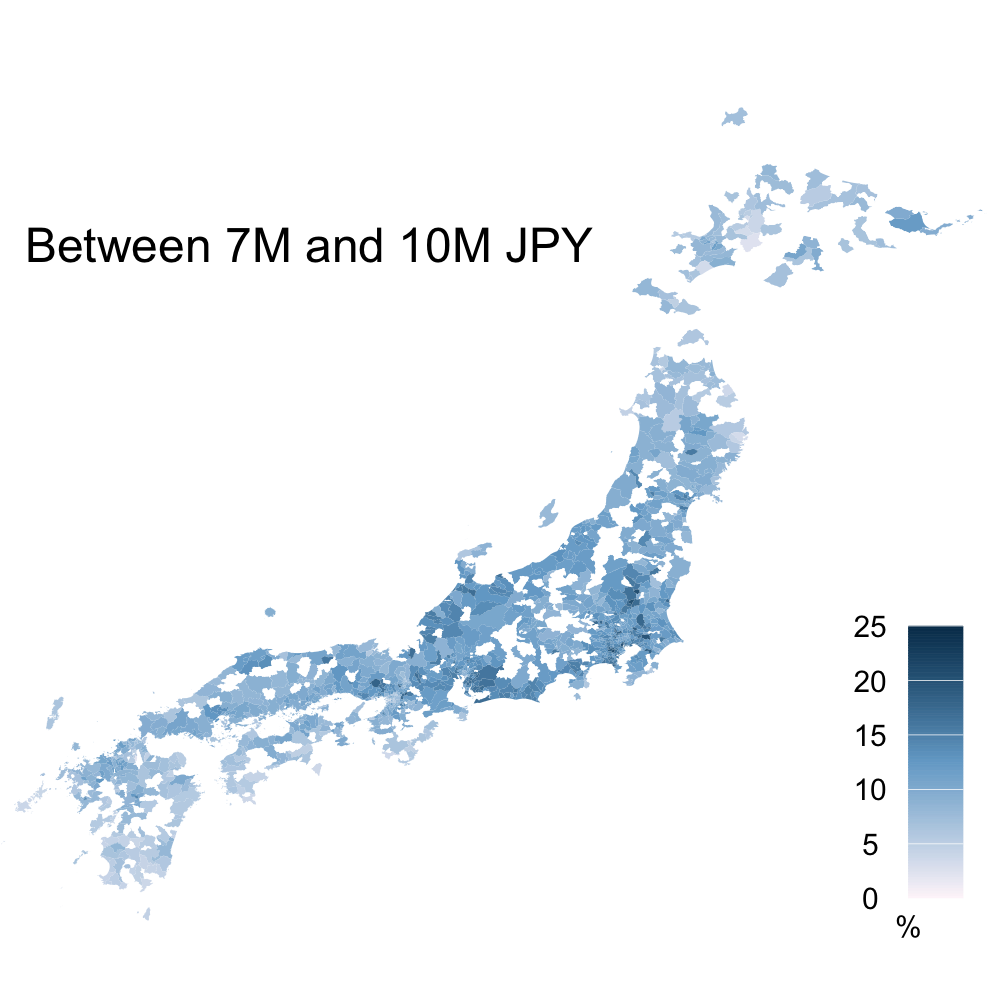} &
\includegraphics[scale=0.14]{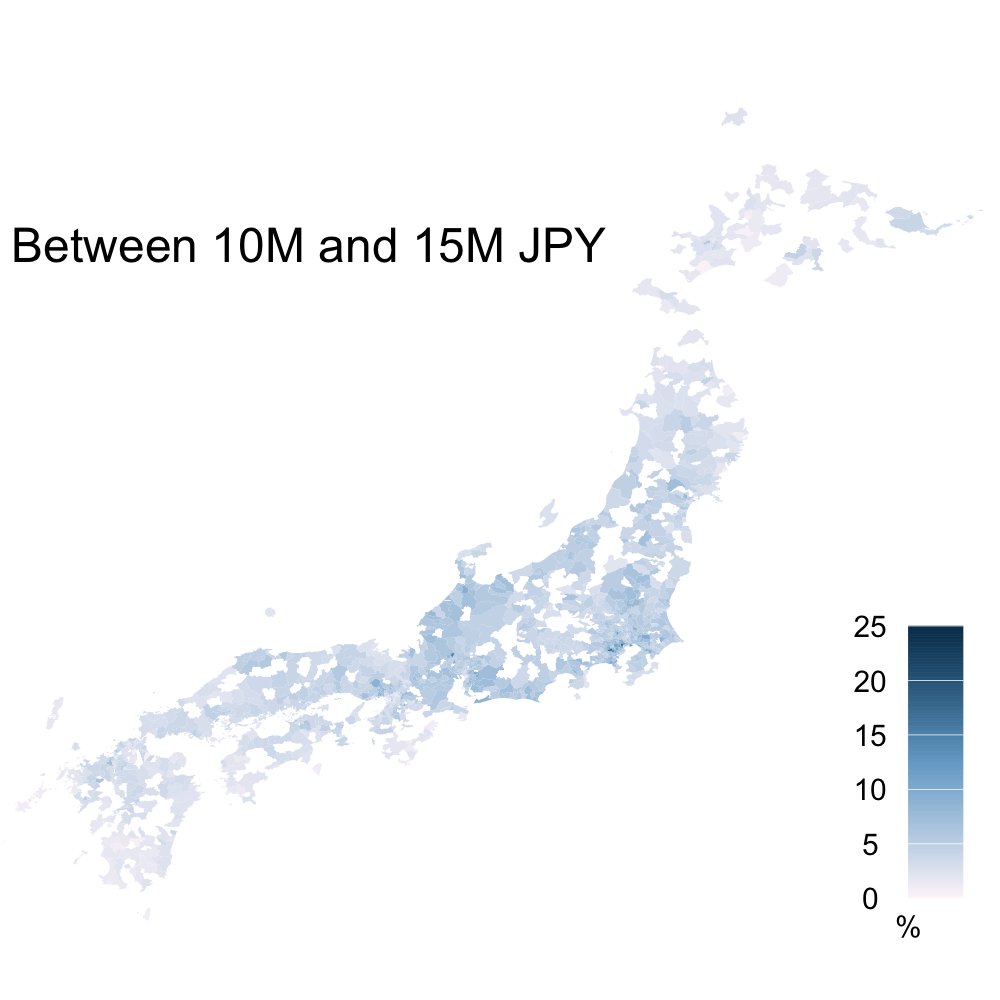} &
\includegraphics[scale=0.14]{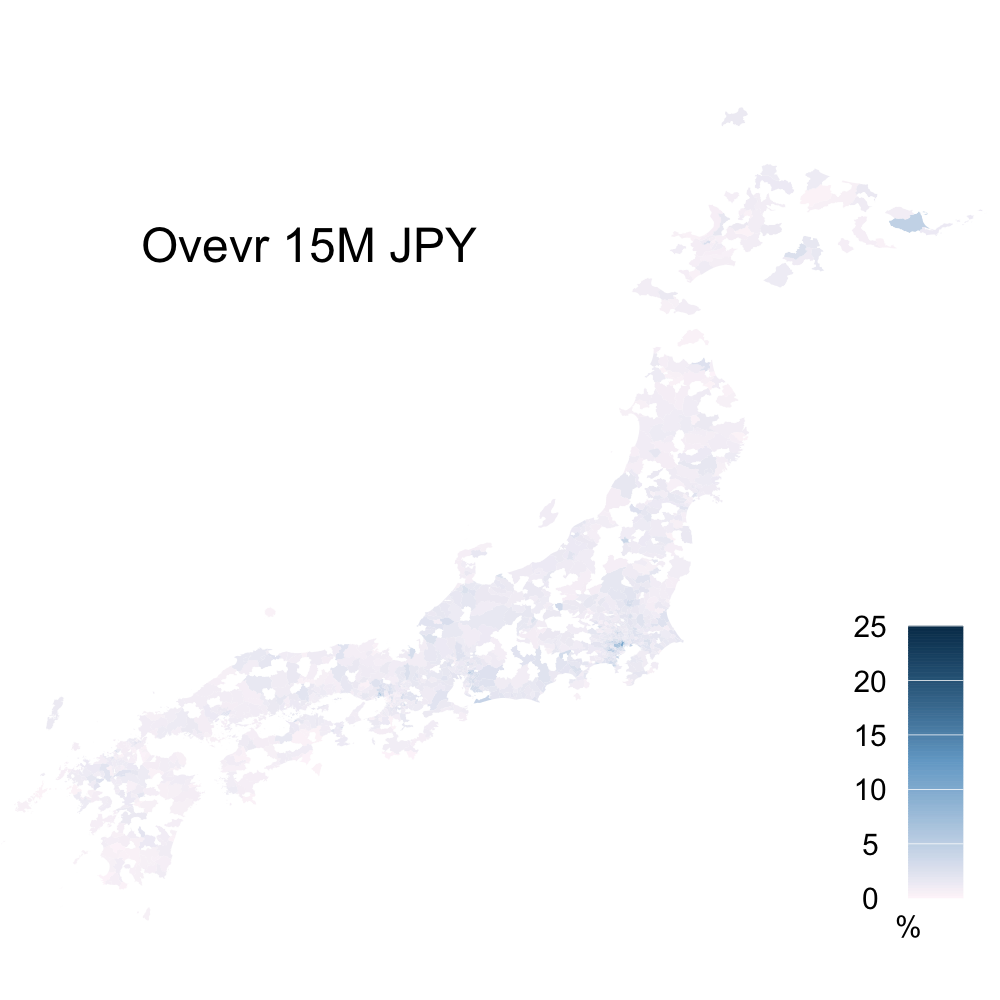} 
\end{tabular}
\caption{Proportions of households in in-sample-municipalities ($G=9$)}
\label{fig:real1}
\end{figure}

The proposed method is demonstrated by using the grouped income data obtained from Housing and Land Survey (HLS) of Japan in 2013. 
The data contains the number of households that fall in $G=5$ and $9$ income classes.\footnote{
Only are the numbers of households in each income class adjusted for the population sizes  accessible in the HLS data and  the original sample sizes for the sampled municipalities of HLS are not published. 
How they are estimated for this analysis is described in Appendix~\ref{sec:app2}.
}
The income classes are defined in million Japanese Yen (M~JPY) and  the thresholds are given  by $(c_1,c_2,c_3,c_4)=(3,5,7,10)$ for $G=5$ and $(c_1,c_2,c_3,c_4,c_5,c_6,c_7,c_8)=(1,2,3,4,5,7,10,15)$ for $G=9$.
In this survey in 2013, 1265 out of 1899 municipalities in Japan were sampled.  
As a summary of the data, Figure~\ref{fig:real1} presents the proportions of the households in the in-sample-municipalities for each income class in the case of $G=9$. 
The maps look incomplete because of the presence of the out-of-sample municipalities. 

Using the proposed method, the EB estimates of the areal mean incomes and Gini coefficients are obtained. 
For the auxiliary variables, we use the total population denoted by $\mathrm{P}_i$ and working-age population denoted by $\WA_i$ obtained from Population Census (PC) of Japan in 2010 and set $\x_i=(1,\log \mathrm{P}_i, \log \WA_i)$ for the $i$th municipality. 
Since these auxiliary variables are also available for the out-of-sample municipalities of HLS, the model can be further utilised to complete the maps of the  mean incomes and Gini coefficients. 


\begin{figure}[H]
\center
\includegraphics[width=15cm]{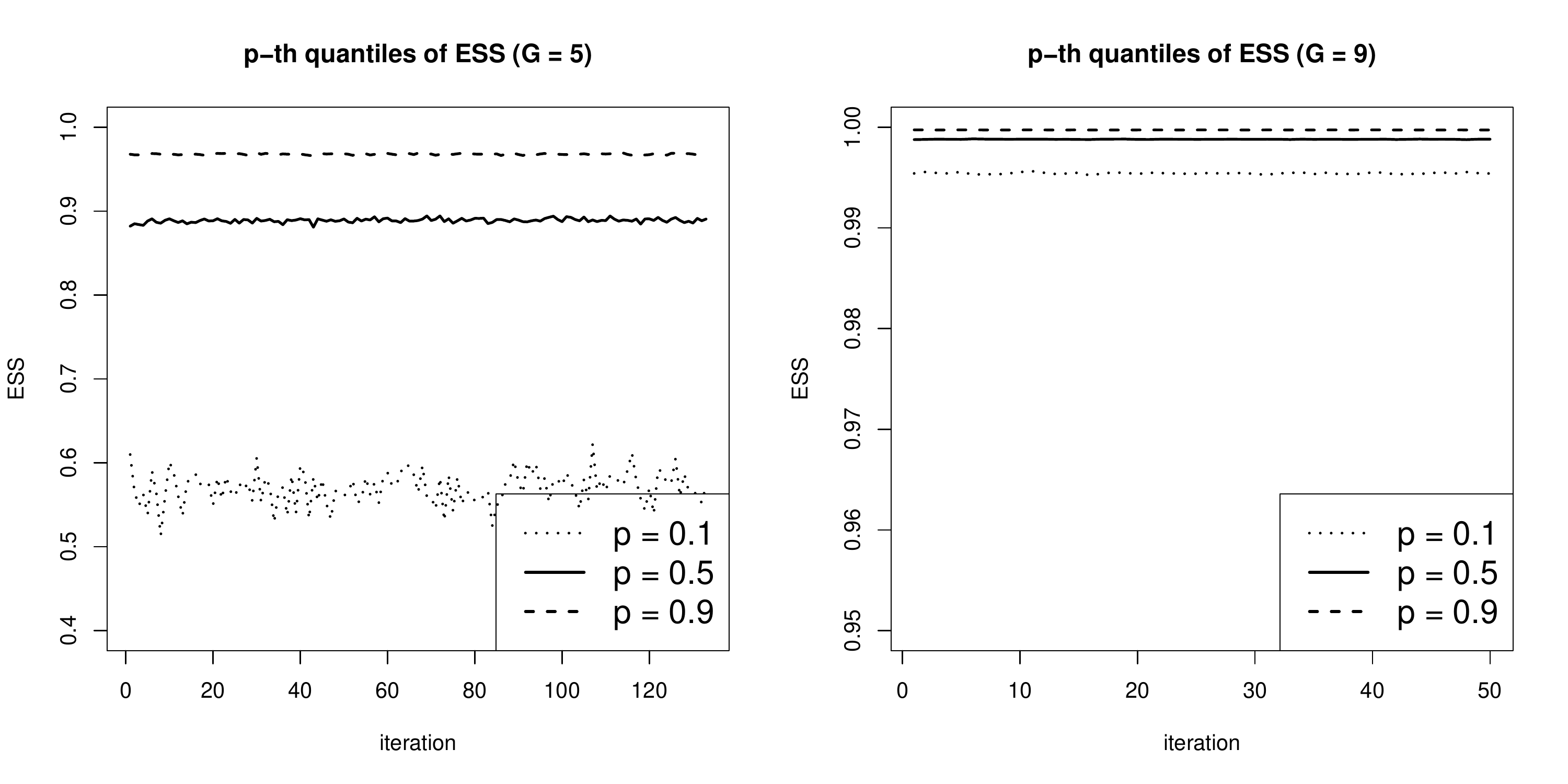}
\caption{Quantiles of effective sample size (ESS) }
\label{fig:real2}
\end{figure}

To estimate the hyperparameters, we set $S_0=100$, $S_1=10000$, $S_2=500$, $H=30$, $d=5$, and $\delta=\epsilon=0.001$ for the MCEM algorithm. 
The initial values are determined using the method described in Section~\ref{subsec:EM}. 
The convergence of the MCEM algorithm occurs relatively fast. 
We also tried other initial values obtained similar results. 
It is noted that the method in Section~\ref{subsec:EM} took much shorter computing times. 
Figure~\ref{fig:real2} presents the $0.1$, $0.5$ and $0.9$ quantiles of the effective sample size (ESS) divided by $S_1$ for the 1265 municipalities at each step of the MCEM algorithm. 
It is seen that the ESS is fairly high and stable over the EM iterations, especially for $G=9$. 

The Bayes estimator of $\zo_i$ is denoted by $\xi_{1i}(\bpsi;\y_i) = E( \zo_i \mid \y_i )$ and that of $\Gini(\z_i)$ is denoted by $\xi_{2i}(\bpsi;\y_i) = E[ \Gini(\z_i) \mid \y_i ]$. 
The EB estimates of $\zo_i$ and $\Gini(\z_i)$ are calculated from the output of the Gibbs sampler \eqref{eqn:full} as
$$
\widehat{ \xi_{1i}(\bpsih;\y_i) } = {1 \over  S_3}\sum_{s=1}^{S_3} \left\{ {1 \over N_i} \sum_{j=1}^{N_i}h_\kah^{-1}(v_{ij}^{(s)}) \right\},
$$
and
$$
\widehat{ \xi_{2i}(\bpsih;\y_i) } = {1 \over S_3}\sum_{s=1}^{S_3} {1 \over N_i} \left\{ N_i + 1 - {2\sum_{j=1}^{N_i}(N_i + 1 - j)h_\kah^{-1}(v_{i(j)}^{(s)}) \over \sum_{j=1}^{N_i}h_\kah^{-1}(v_{ij}^{(s)}) } \right\},
$$
where $\{ v_{i(1)}^{(s)},\dots,v_{i(N_i)}^{(s)} \}$ are sorted values of $\{ v_{i1}^{(s)},\dots,v_{iN_i}^{(s)} \}$ in non-decreasing order. 
In this analysis, we run the Gibbs sampler for $S_3=500$ iterations with the initial burn-in period of $50$ iterations.

While it is generally difficult to define a reasonable direct estimator for these small area parameters from grouped data, for a comparison purpose,  we may also think of the following ``naive" estimator of the areal mean $\zo_i$ that uses the class midpoints given by
\begin{equation}
\label{eqn:naive}
\widehat{\zo}_i^{\mathrm{naive}} = {1 \over n_i} \sum_{g=1}^G \co_g \times y_{ig}
\end{equation}
where $\co_g = ( c_{g-1} + c_g ) / 2$ for $g=1,\dots,G-1$ and $\co_G = c_{G-1} + ( c_{G-1} - c_{G-2} ) / 2$. 
This estimator is naive particularly because the upper end $\co_G$ has to be set and its choice is completely arbitrary. 
The choice of $\co_G$ would have a huge impact on its performance. 
Note that the proposed approach has no arbitrariness with this respect as $c_G=\infty$ and \eqref{eqn:prob_g} is well defined.  

\begin{figure}[H]
\center
\begin{tabular}{cc}
\includegraphics[width=7.5cm]{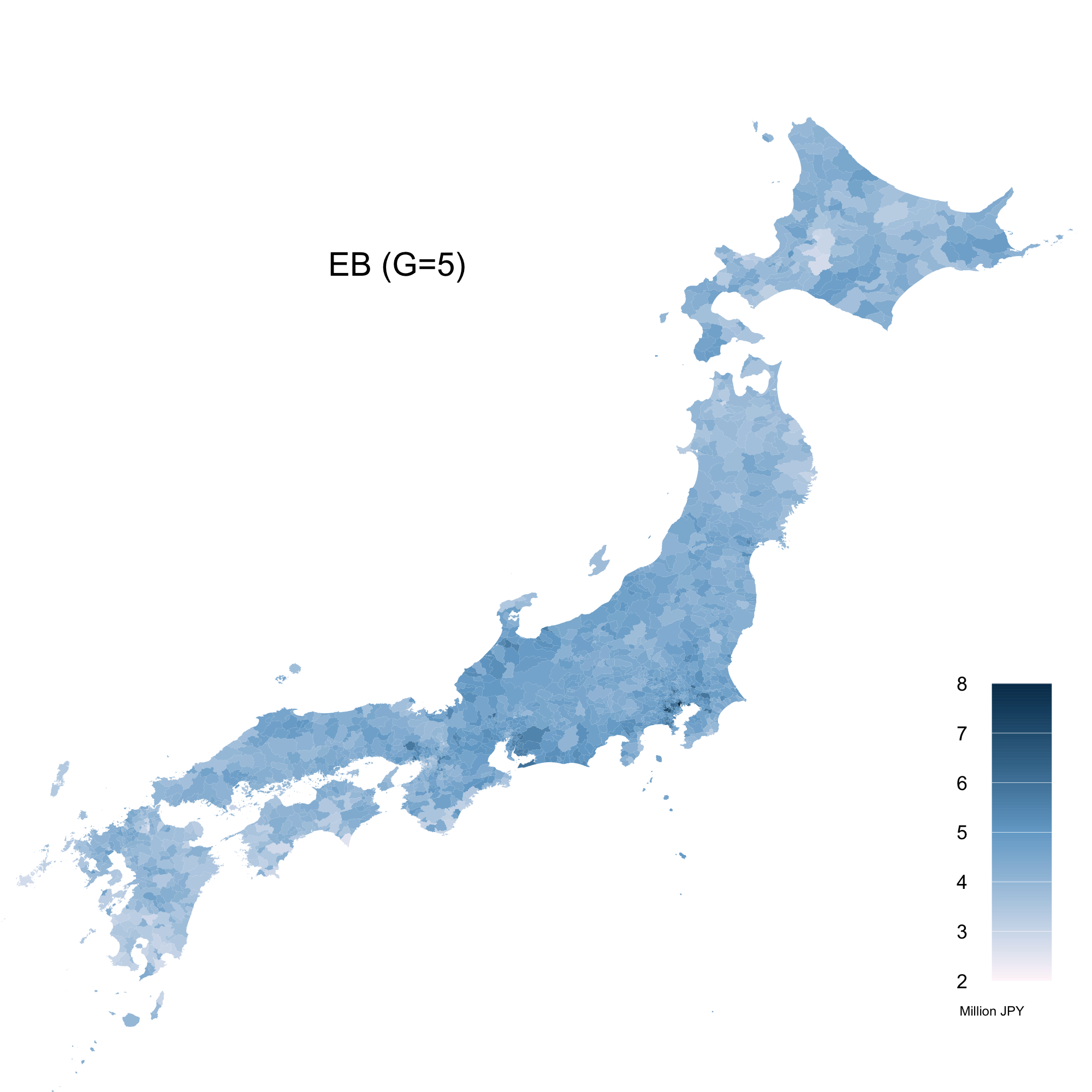} &
\includegraphics[width=7.5cm]{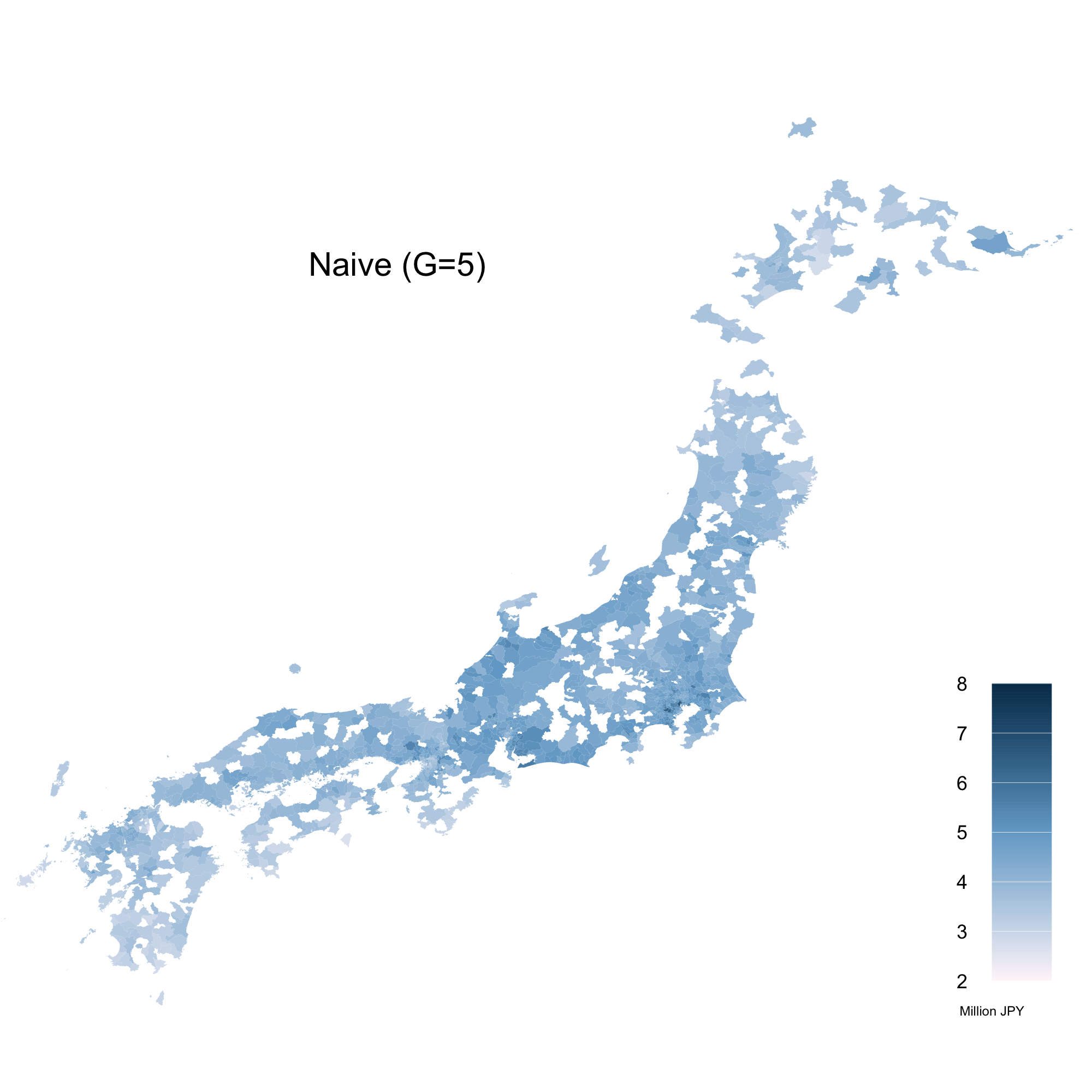} \\
\includegraphics[width=7.5cm]{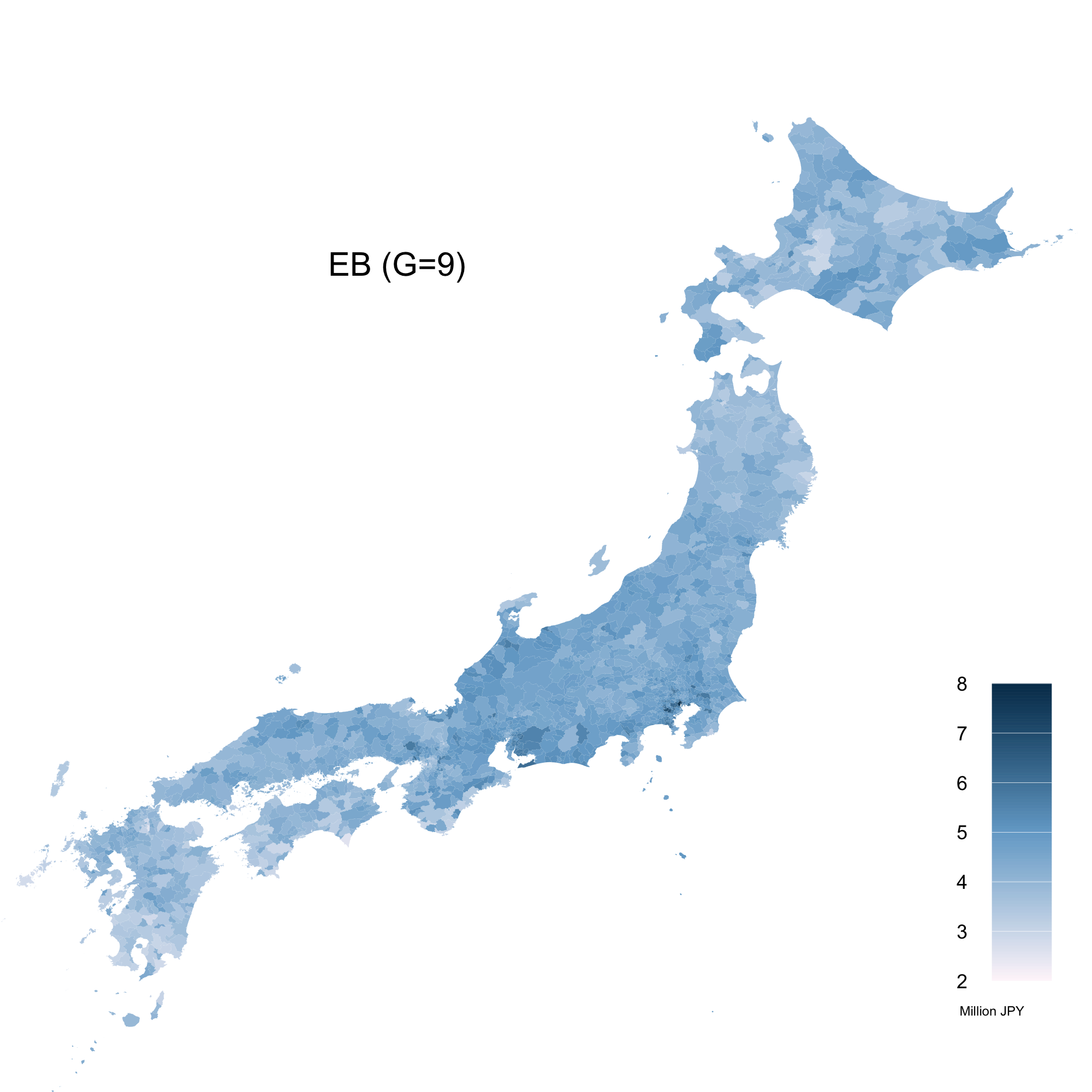} & 
\includegraphics[width=7.5cm]{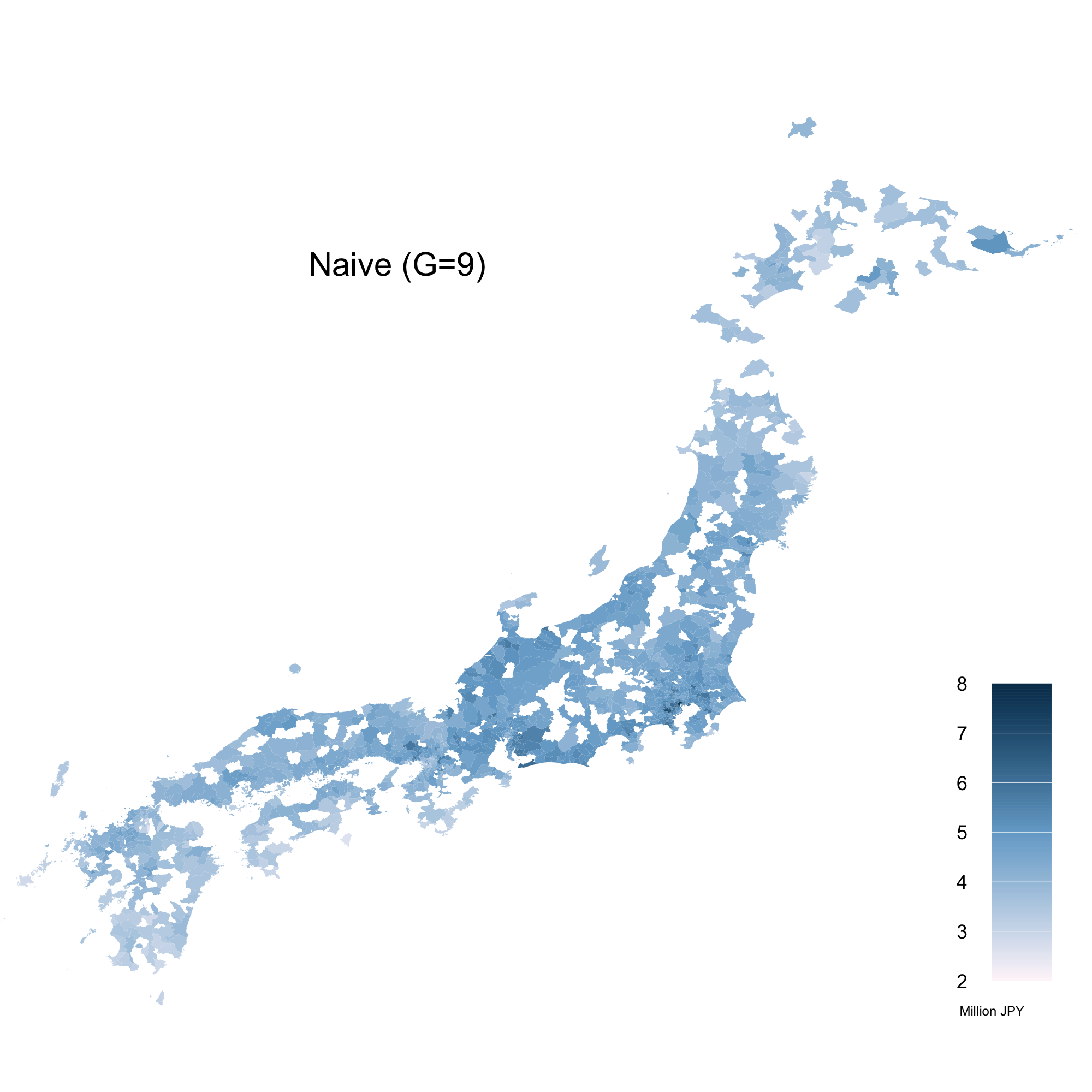} \\
\end{tabular}
\caption{EB and naive estimates of areal means}
\label{fig:real_mean}
\end{figure}

\begin{figure}[H]
\center
\includegraphics[scale=0.65]{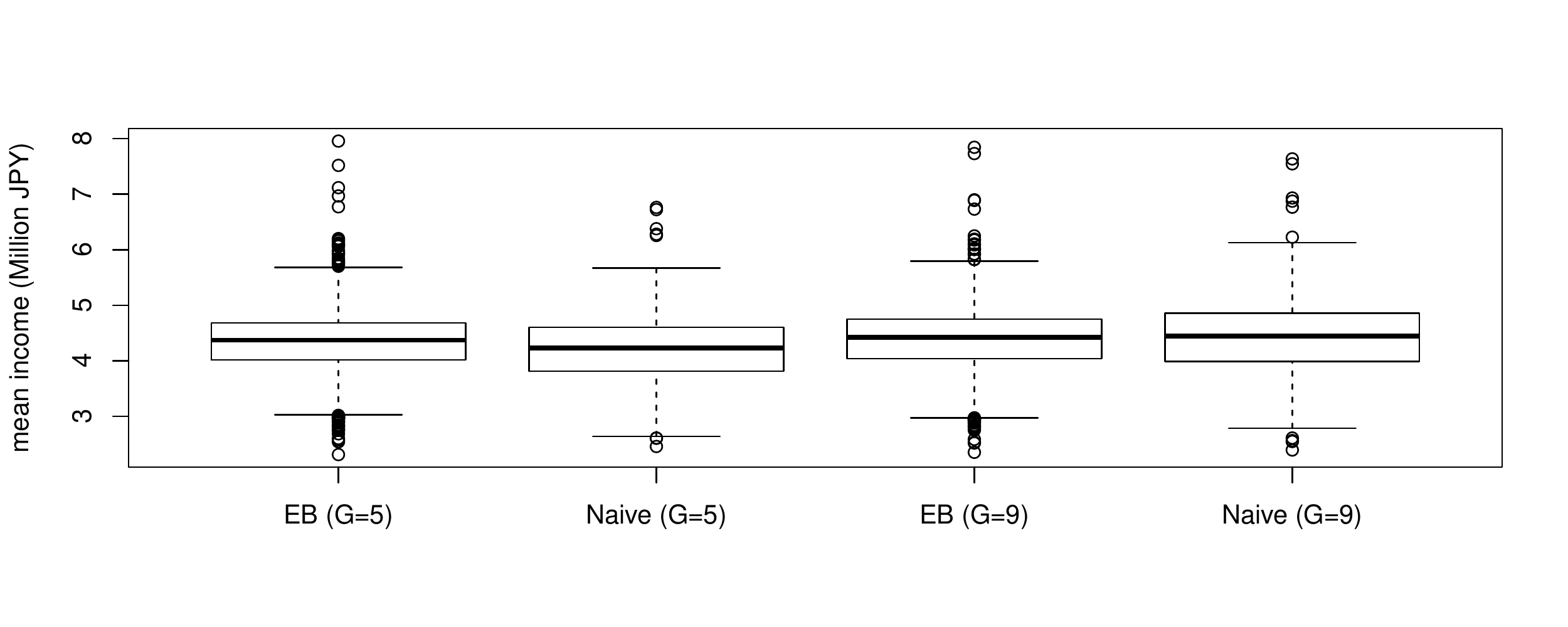}
\caption{Boxplots of EB and naive estimates of areal means for the sampled areas}
\label{fig:real_Box}
\end{figure}

\begin{figure}[H]
\center
\includegraphics[scale=0.65]{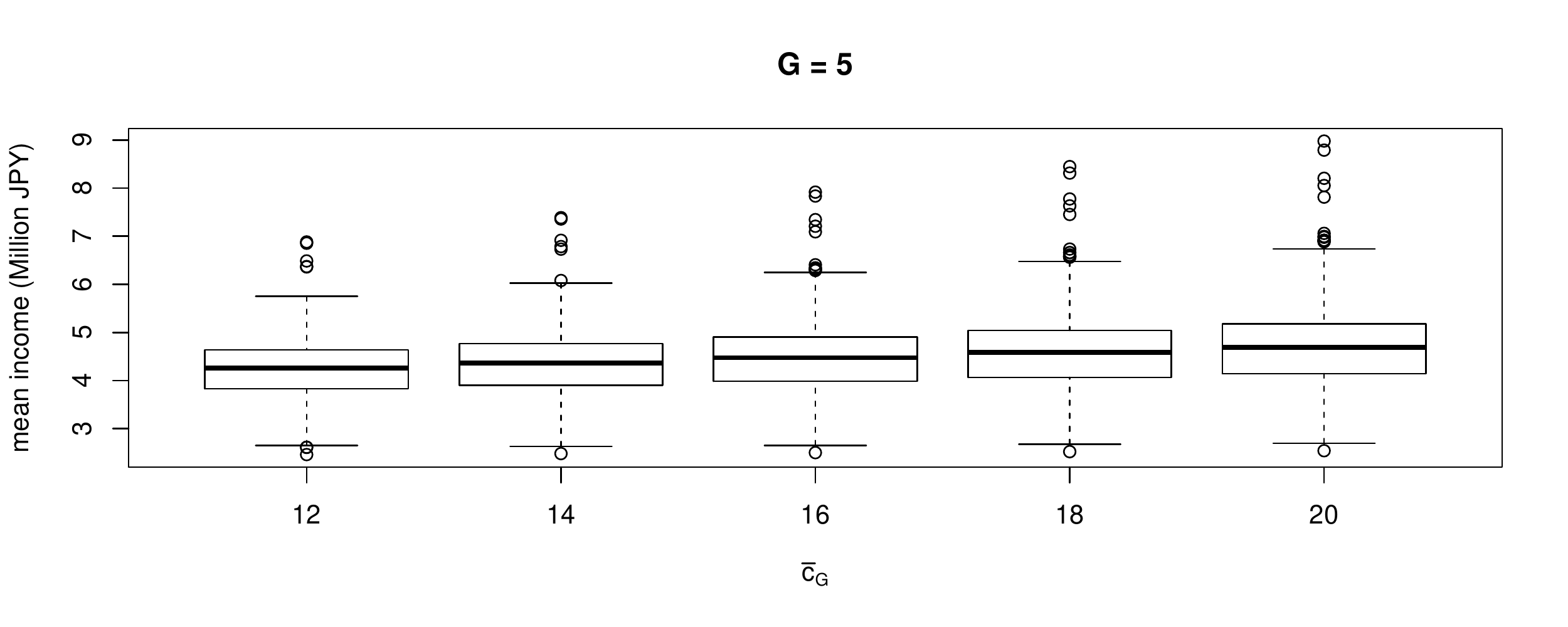}\\
\includegraphics[scale=0.65]{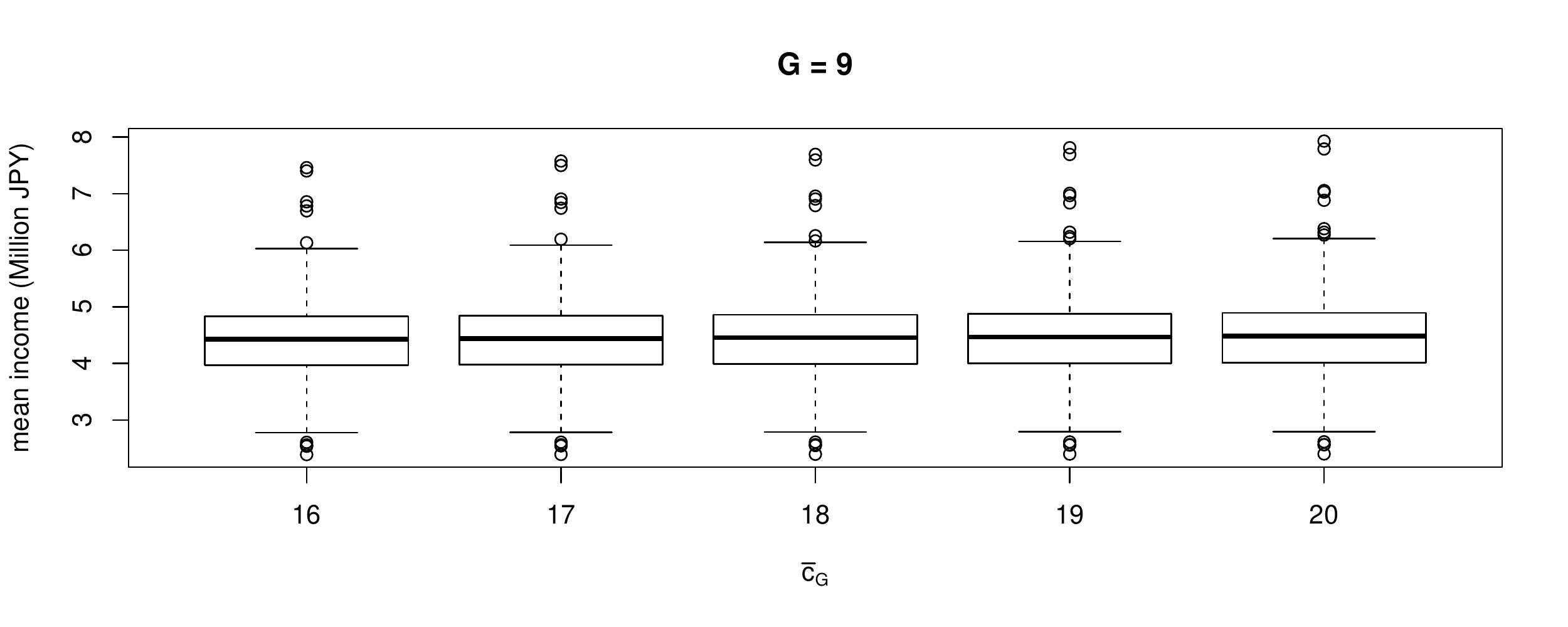}
\caption{Boxplots of naive estimates of areal means under different values of $\bar{c}_G$}
\label{fig:real_Box_d}
\end{figure}

Figure~\ref{fig:real_mean} presents the estimates of the areal means based on the proposed method and naive method \eqref{eqn:naive}.
By borrowing strength from the other municipalities through the statistical model \eqref{eqn:lmm}, the proposed method can predict the income for the out-of-sample municipalities and provide the complete maps of the mean incomes and Gini coefficients. 
The boxplots of Figure~\ref{fig:real_Box}  compares the EB and naive estimates of the areal means for the sample areas. 
The figure indicates that the results for the naive estimates can vary between $G=5$ and $9$ resulting the lower mean incomes for some areas for $G=5$ than for $G=9$.
This would be because the naive estimates cannot capture the behavior of the upper tail of the income distribution, which has an impact on the estimation of the mean income. 
In fact, we also considered the different values for $\bar{c}_G$ for the naive estimates to demonstrate the impact. 
Figure~\ref{fig:real_Box_d} presents the boxplots of the naive estimates under the different values of $\bar{c}_G$ for $G=5$ and $9$. 
The figure shows that the naive estimates exhibit severe sensitivity with respect to the setting of $\bar{c}_G$ in the case of  $G=5$. 
While the sensitivity decreases for $G=9$, the areal mean estimates for the high income areas still appear to increase with $\bar{c}_G$. 

\begin{figure}[H]
\center
\begin{tabular}{cc}
\includegraphics[width=7.5cm]{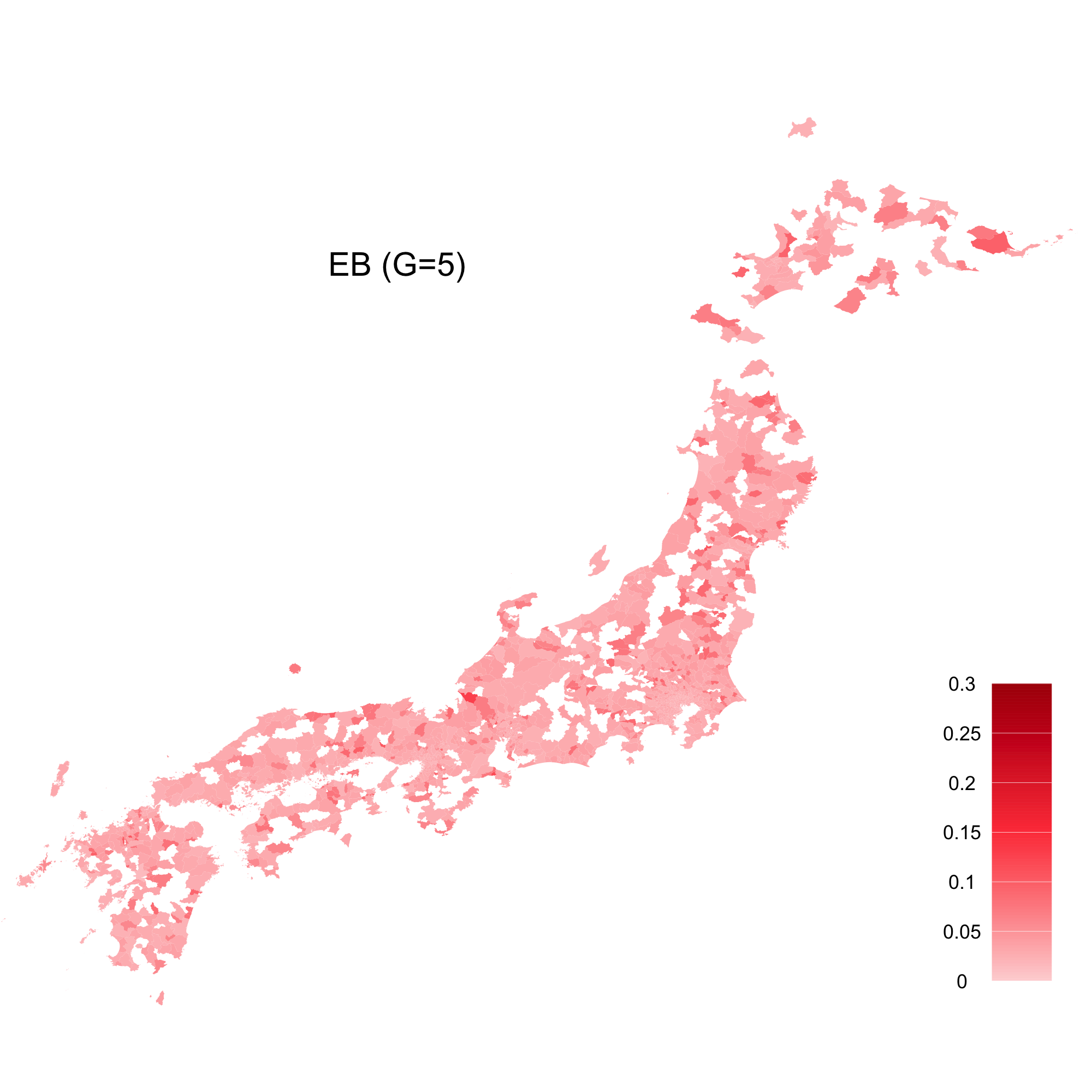}&
\includegraphics[width=7.5cm]{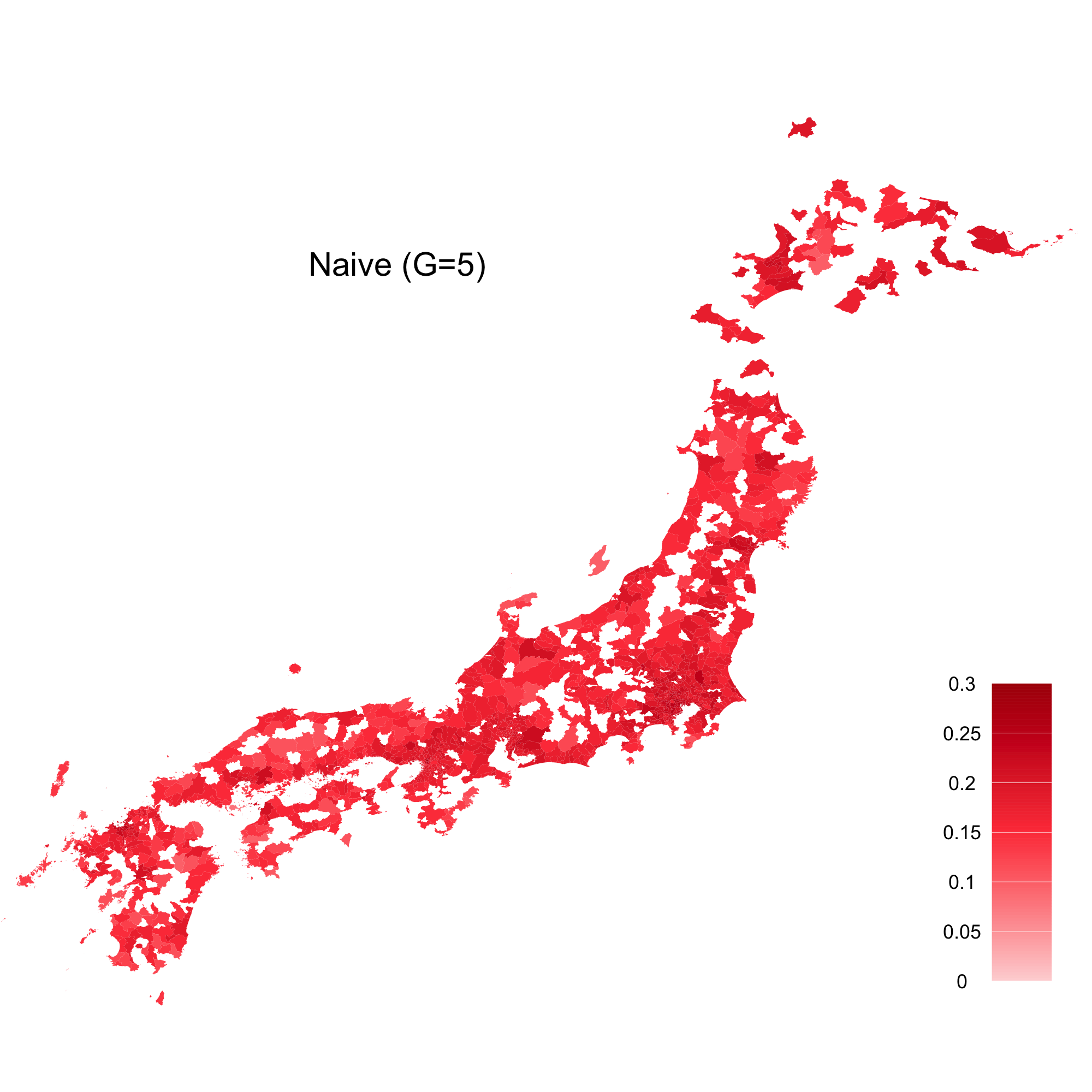}\\
\includegraphics[width=7.5cm]{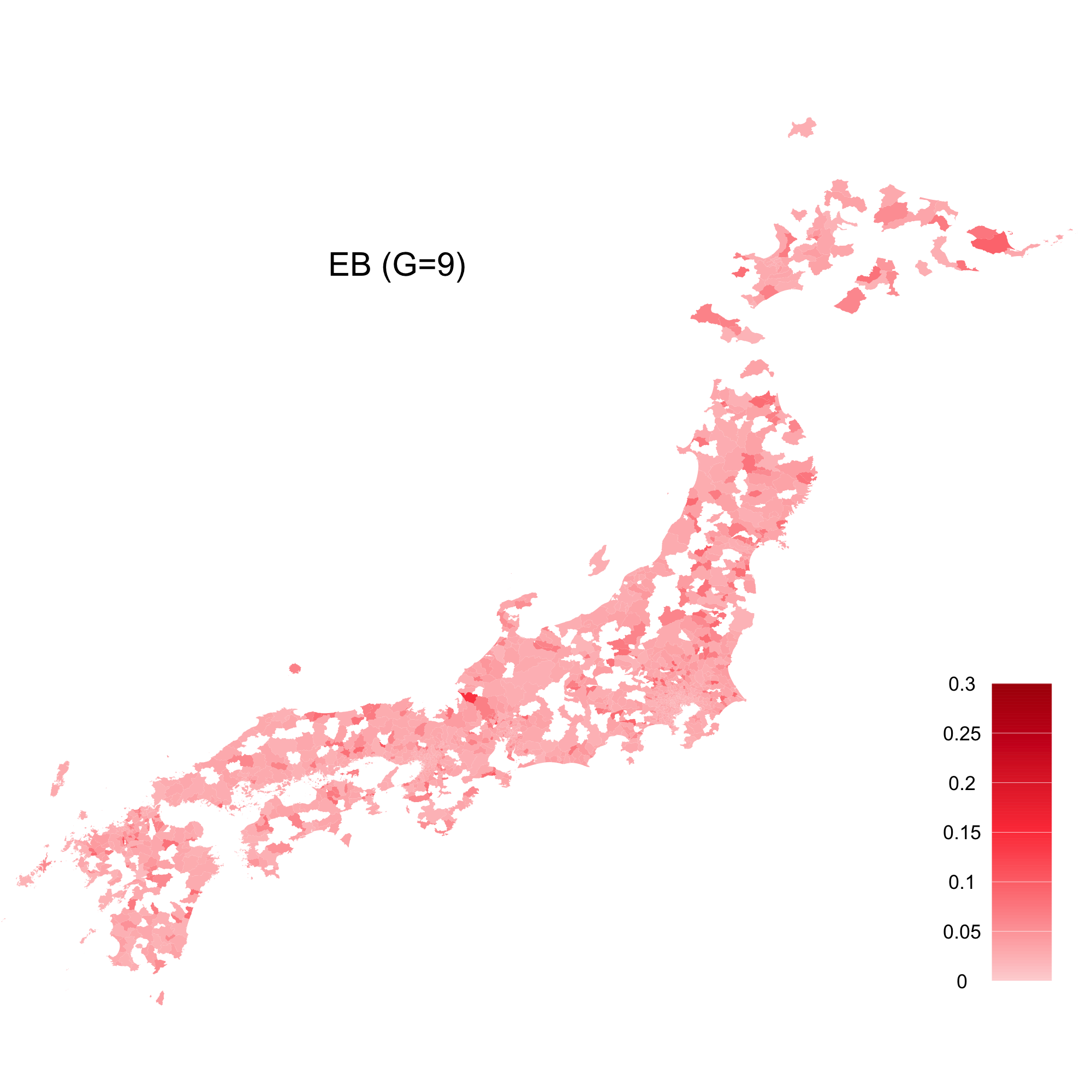}&
\includegraphics[width=7.5cm]{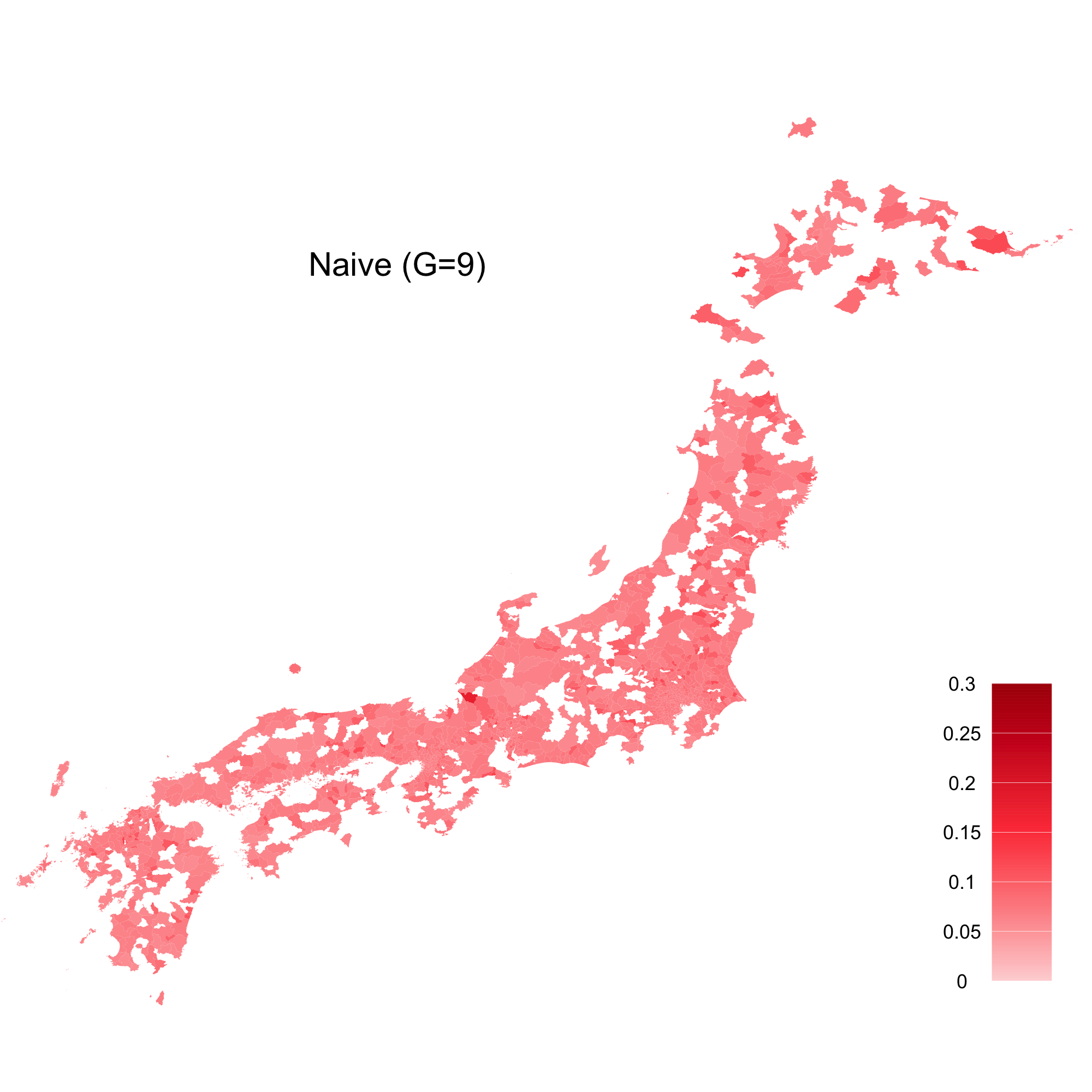}\\
\end{tabular}
\caption{Estimates of RMSE of the naive estimators and EB estimators for areal means}
\label{fig:real_RMSE}
\end{figure}

In order to assess the uncertainty of the estimators, we estimated the root mean squared error (RMSE) of the estimators for the sampled municipalities by using a parametric bootstrap method.
Let $z_{ij}^{*(b)} \ (i=1,\dots,m; \ j=1,\dots,N_i)$ and $\{ \y_1^{*(b)}, \dots, \y_m^{*(b)} \}$ denote the $b$th bootstrap sample  $(b=1,\dots, B)$ generated from the models (\ref{eqn:model_y}) and (\ref{eqn:lmm}) with the hyperparameter fixed to the maximum likelihood estimate  $\bpsih$.
Then, the RMSE of the EB estimator of areal mean is estimated as
\begin{equation*}
\widehat{\rm RMSE}_i^{\rm EB} = \sqrt{ {1 \over B}\sum_{b=1}^B \left\{ \widehat{ \xi_{1i}(\bpsih; \y_i^{*(b)}) } - \zo_i^{*(b)} \right\}^2 },
\end{equation*}
for a large $B$, where $\zo_i^{*(b)} = N_i^{-1}\sum_{j=1}^{N_i}z_{ij}^{*(b)}$.
For each $b$, we simply run the Gibbs sampler described in Section~\ref{subsec:Gibbs} to calculate the EB estimates given the estimate $\bpsih$ from the original data, not on the bootstrap samples.
In the same way, the RMSE of the naive estimator is estimated as
$$
\widehat{\rm RMSE}_i^{\rm naive} = \sqrt{ {1 \over B}\sum_{b=1}^B \left\{ \widehat{\zo}_i^{\mathrm{naive}*(b)} - \zo_i^{*(b)} \right\}^2 },
$$
where $\widehat{\zo}_i^{\mathrm{naive}*(b)} = n_i^{-1}\sum_{g=1}^G \co_g \times y_{ig}^{*(b)}$.
Figure~\ref{fig:real_RMSE} presents the estimates of the RMSE of the EB estimators and naive estimators for the sampled areas.
The naive estimators resulted in the large RMSE indicated by the darker shade of red in the case of $G=5$. 
While the RMSE for the naive estimators improves as the number of income classes increases, the EB estimators resulted in the smaller RMSE.
The figure also shows that the overall improvement in the RMSE of the EB estimators in the case of $G=9$ over $G=5$  is marginal compared to the naive estimators.

\begin{figure}[H]
\center
\begin{tabular}{cc}
\includegraphics[width=7.5cm]{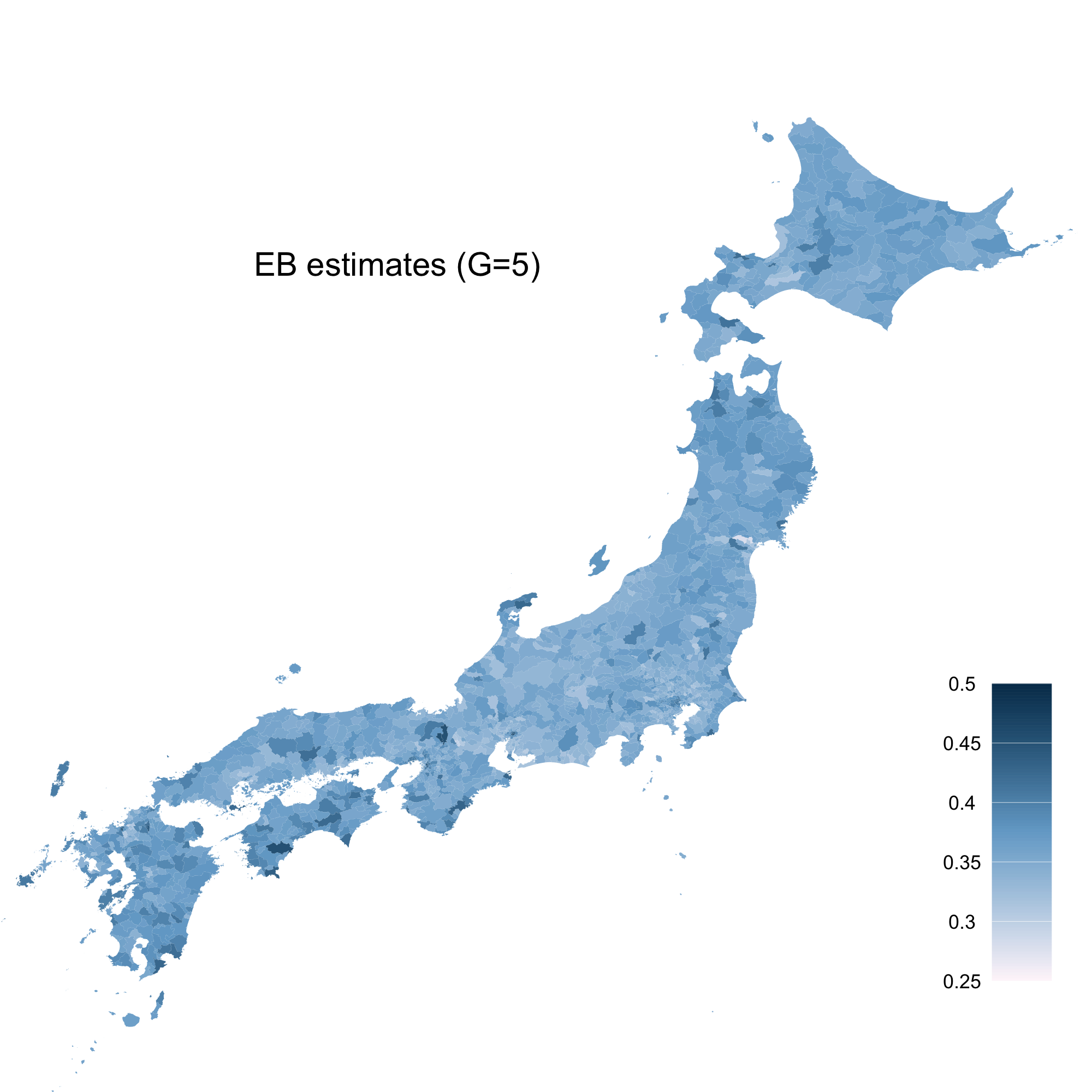}& 
\includegraphics[width=7.5cm]{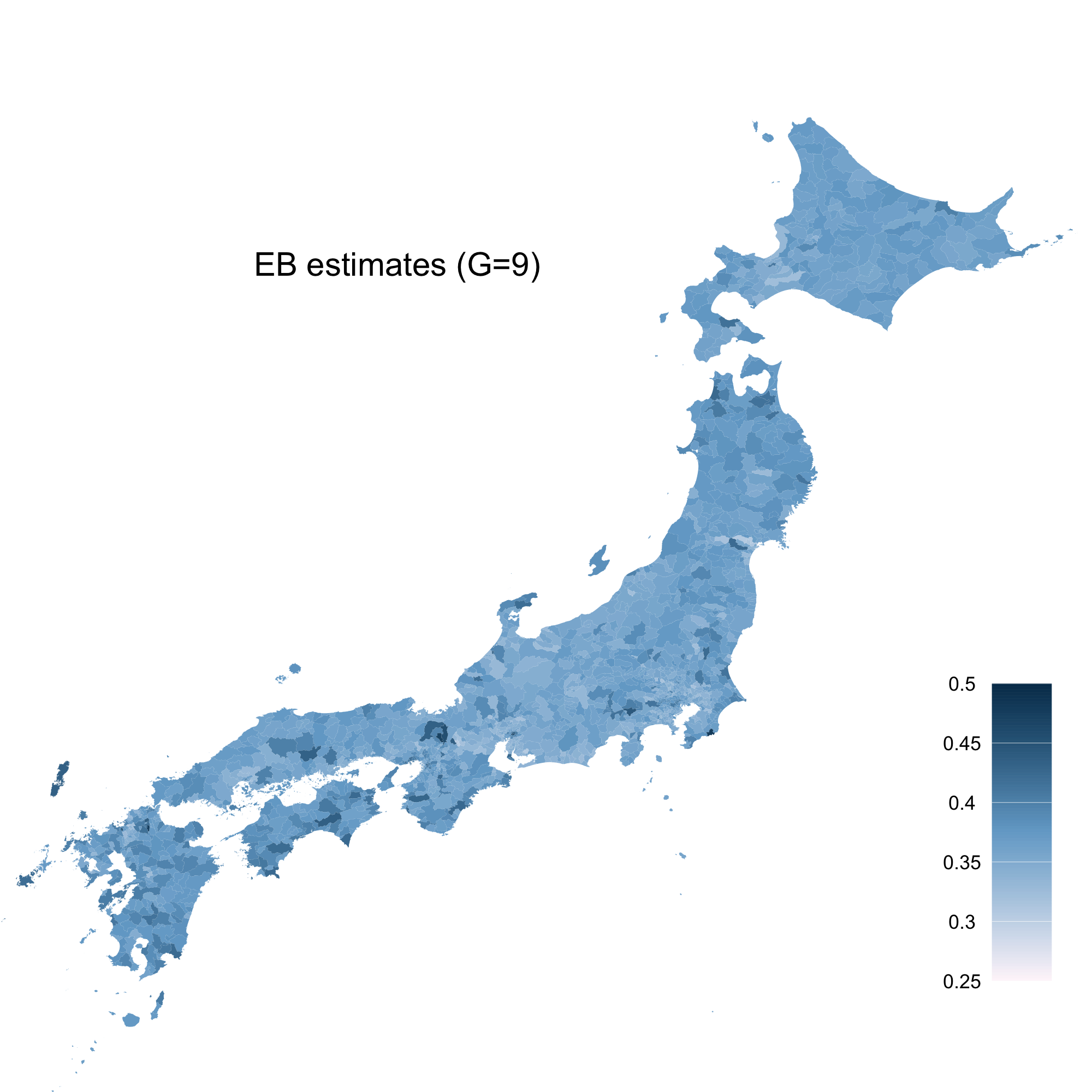}\\
\includegraphics[width=7.5cm]{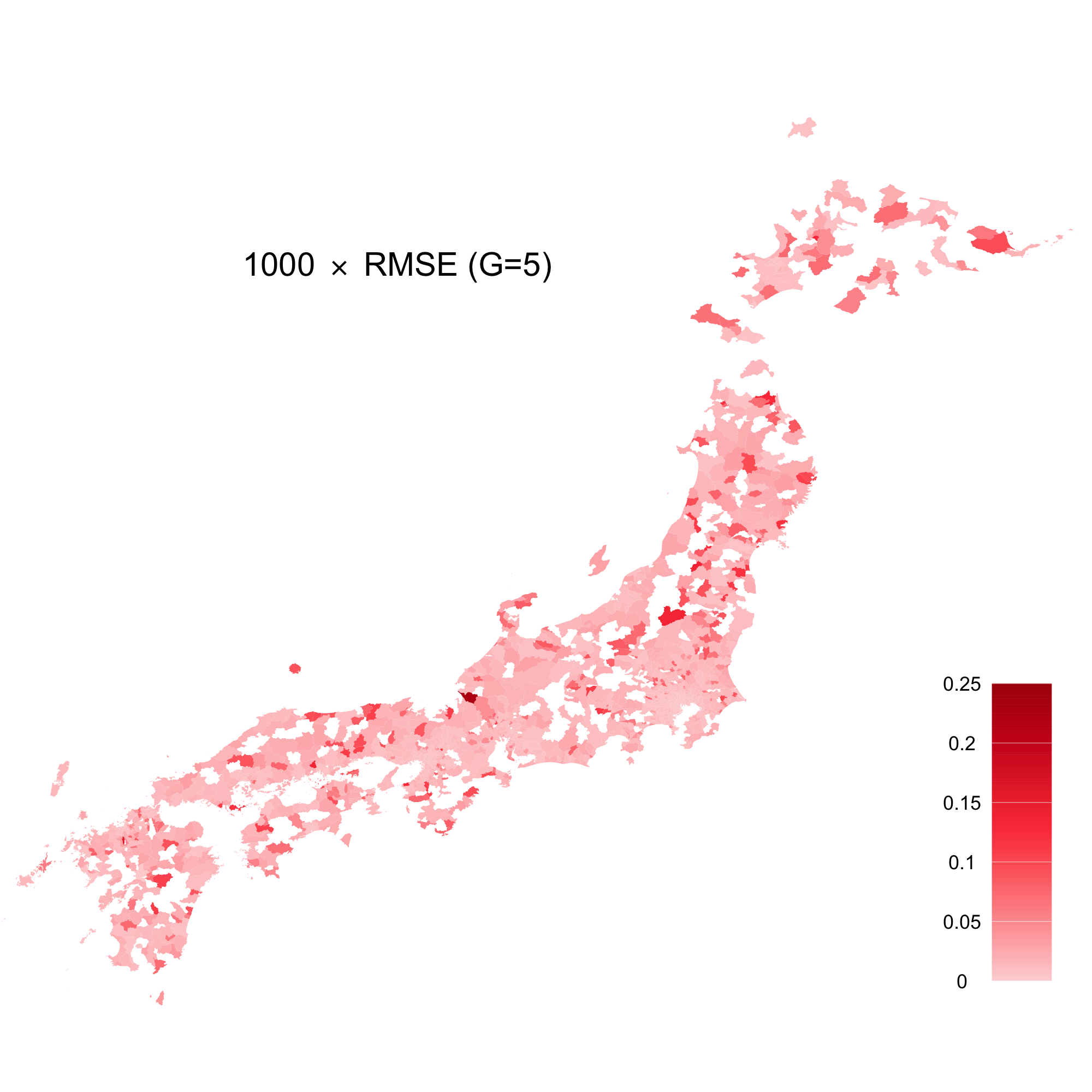}&
\includegraphics[width=7.5cm]{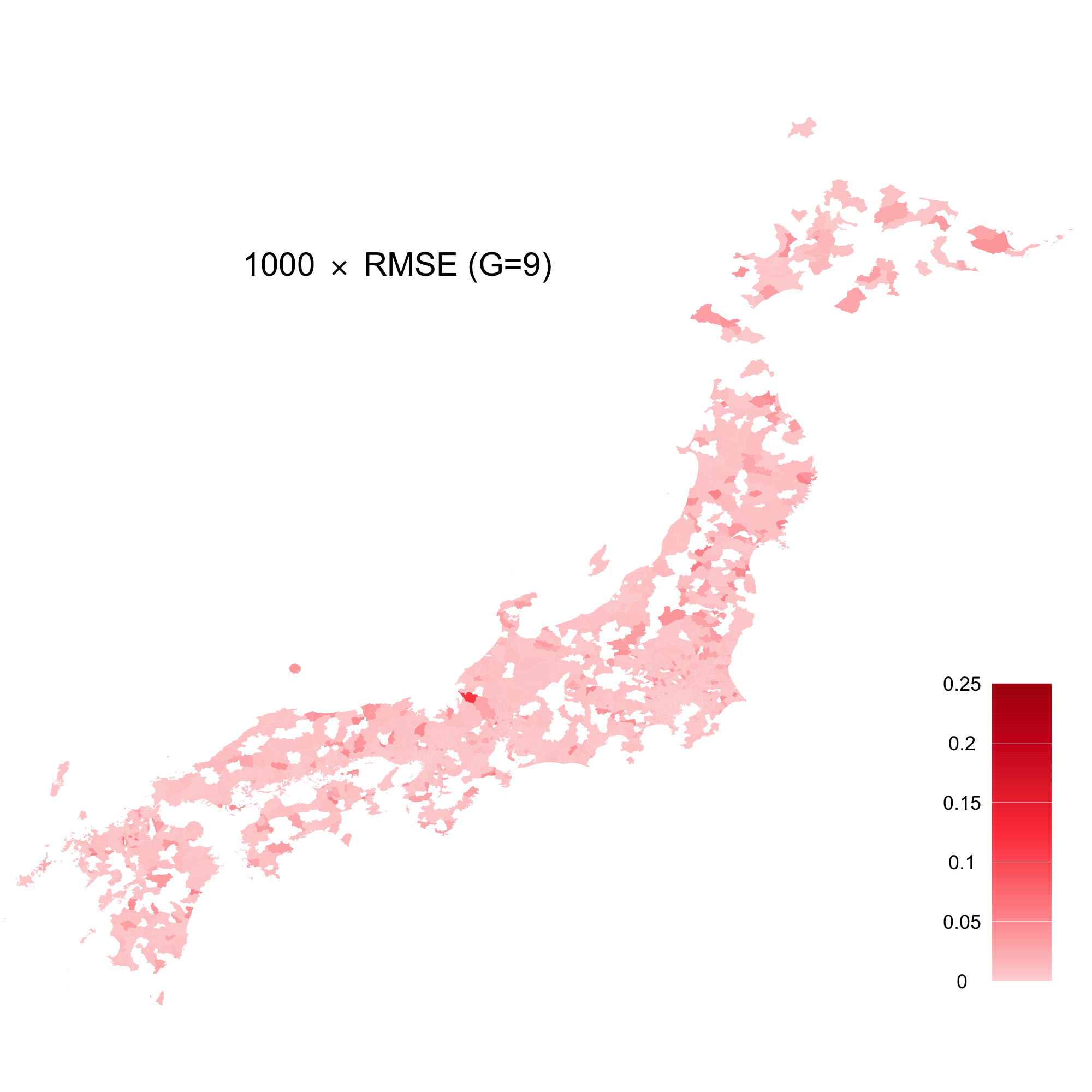}
\end{tabular}
\caption{EB estimates and estimates of RMSE (multiplied by $1000$) for Gini coefficients}
\label{fig:real_Gini}
\end{figure}

Finally, Figure~\ref{fig:real_Gini} presents the EB estimates for the Gini coefficients for all municipalities and associated estimates of RMSE for the sampled municipalities. 
As in the case of the mean incomes, the proposed method can also predict the Gini coefficients for the out-of-sample municipalities to complete the map.
The RMSE of the estimator of the Gini coefficient is estimated in the same way as that of the mean income by using the parametric bootstrap.
The map for the case of $G=9$ exhibits darker shades of blue than the map for $G=5$ implying that the degree of inequality is greater across the country. 
This could be because that the data with $G=9$ contains more information on the income distribution, especially on the upper tail of the distribution which can have an impact on the estimates. 
The figure also shows that the uncertainty regarding the Gini coefficients estimation decreases as the number of income classes in the data increases.

\section{Simulation Studies}
\label{sec:sim}

\subsection{Model-based simulation}
In this section, the proposed approach is illustrated using the simulated data.
The first simulation is a model-based simulation where \eqref{eqn:lmm} is the  data generating process. 
The true parameter values are set to the estimates obtained in the real application in Section \ref{sec:income} and we use the same values of the auxiliary variables $\x_i$'s as the real data for the randomly chosen $m=100$ areas out of the 1265 in-sample areas  of HLS.
Based on this setting, we generate $R=100$ replications of $z_{ij}$'s with $N_i = 1000$ for all $i$ and calculate the true mean $\zo_i$ and Gini coefficient $\mathrm{GINI}(\z_i)$.
For each replication, we obtain a  frequency distribution for each area from  the simulated data $\{ z_{i1},\dots,z_{i,n_i} \}$. 
The two cases of the numbers of groups  $G=5$ and $9$ with the same thresholds as HLS are considered. 
The sample sizes are set as $n_i = 10 \ (i=1,\dots,20), \ n_i = 50 \ (i=21,\dots,40), \ n_i = 100 \ (i=41,\dots,60), \ n_i = 150 \ ( i=61,\dots,80 )$, and $n_i = 200 \ (i=81,\dots,100)$.
The true parameter values and the auxiliary variables $\x_i$'s for $i=1,\dots, m$ are fixed for all replications.
The settings for the MCEM algorithm and the Gibbs sampler are the same as the real data analysis in Section \ref{sec:income}.

In order to demonstrate the advantage of the present approach, the naive estimator of $\widehat{\zo}_i^\mathrm{naive}$ in \eqref{eqn:naive} is also considered again.
The performance of the methods is compared by the simulated relative root MSE (RRMSE) over $R=100$ replications of the data. 
The simulated RRMSE is calculated as
\begin{equation*}
{\rm RRMSE}(\widehat{\zo}_i) = \sqrt{{1 \over R}\sum_{r=1}^R \left( { \widehat{\zo}_i^{(r)} - \zo_i^{(r)} \over \zo_i^{(r)} } \right)^2},
\end{equation*}
where $\widehat{\zo}_i^{(r)}$ is the EB or naive estimates and $\zo_i^{(r)}$ is the true mean in the $r$th replication.

\begin{figure}[H]
\center
\includegraphics[width=9cm]{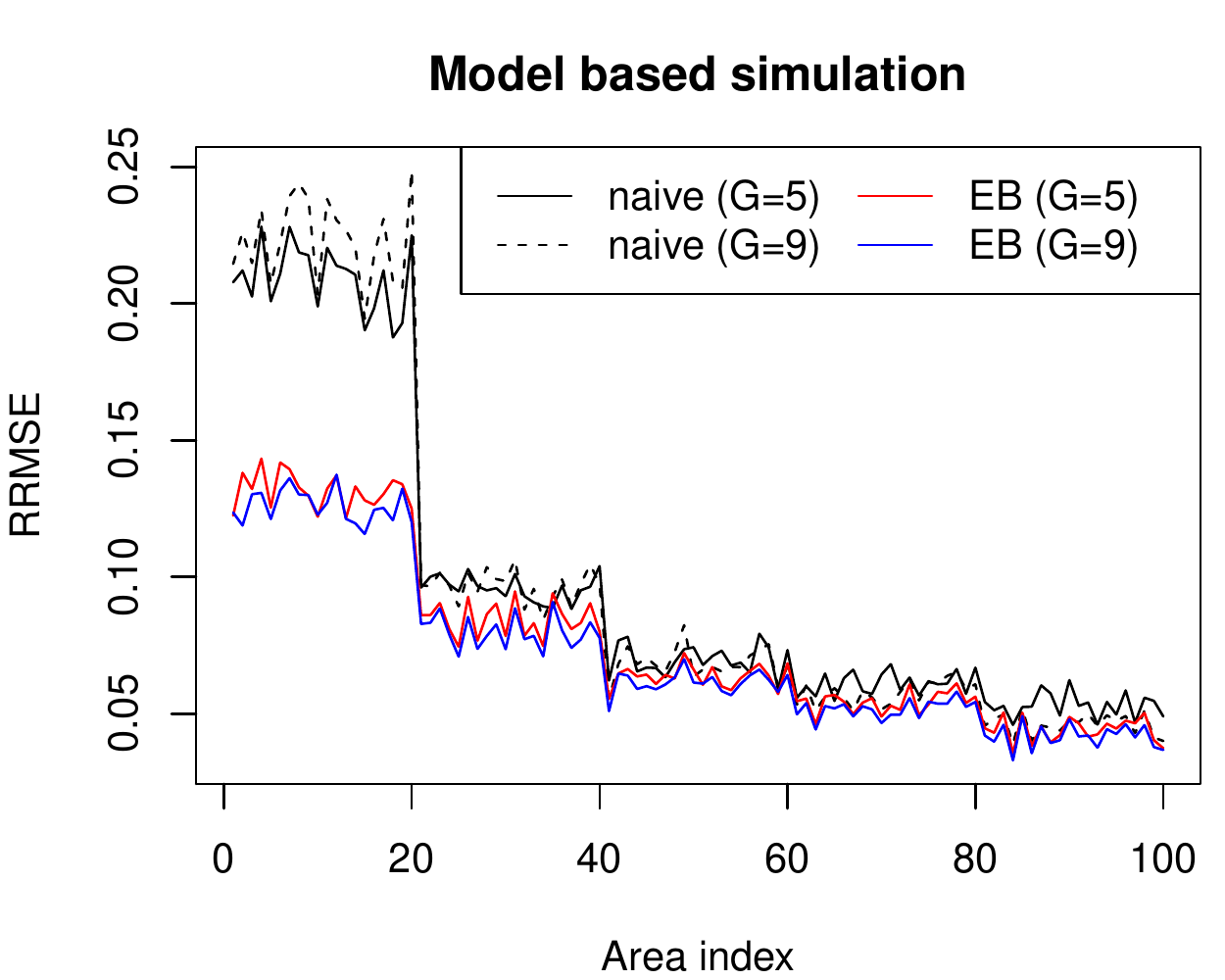}
\caption{RRMSE of EB estimator and naive estimator for model based simulation}
\label{fig:sim1}
\end{figure}

Figure~\ref{fig:sim1} shows the result of the simulation.
Noting that the horizontal axis represents the area index, the figure shows that the RRMSE decreases as the sample size increases both for the EB estimator and the naive estimator.
In terms of RRMSE, the EB estimator improves on the naive estimator for all the areas.
It is interesting to see that the improvement of the RRMSE is much larger for the areas with small sample sizes, especially for the areas with $n_i=10$ and $50$.
This is because the EB estimator borrows strength from other areas even though the area sample size is small, while the naive estimator only uses the information of the target area.
It is also observed that EB estimator for $G=9$ resulted in better performance than  for $G=5$ for most of the areas.
This is a natural result because the frequency distributions based on  $G=9$ contain more information of the distribution of the latent  $z_{ij}$'s.

\subsection{Design-based simulation}
The second simulation is a design based simulation where \eqref{eqn:lmm} is not assumed to be the data generating process. 
For this simulation, the Spanish income dataset included in the R package \texttt{sae} developed by \citet{MM18}. 

This dataset contains the synthetic data on income and some related information of 17199 households including the province where the  household is located and the gender of the head of the household. 
There are 52 provinces in Spain and for each province the dataset is divided based on the gender of the head of the household. 
Therefore, this dataset consists of $m=104$ small domains. 

We generate the datasets for this design-based simulation study following the technique used by \citet{CSCT12}.  
First, a synthetic population is created for each domain by resampling with replacement from the original dataset and calculate the `true' population mean for each dataset. 
Then 100 independent samples are obtained from the fixed synthetic populations based on the simple random sampling without replacement and form a frequency distribution for each domain. 

As the auxiliary variables, we use $\x_i = (1, \mathrm{NAT}_i, \mathrm{WA}_i, \mathrm{LABOR}_i)^\top$ where $\mathrm{NAT}_i$ is the proportion of the people holding Spanish nationality in the $i$th domain, $\mathrm{WA}_i$ is the proportion of the people who are in working age in the $i$th domain, and $\mathrm{LABOR}_i$ is the proportion of the people who are employed in the $i$th domain.
For the transformation in \eqref{eqn:lmm}, since the negative income observations are present for some households in this dataset, the following modified Box--Cox transformation is used:
$$
h_\kappa(z) = { (z-C)^\kappa - 1 \over \kappa },
$$
where $C$ is equal to 0.1 less than the minimum income of the synthetic population. 
The same settings for the MCEM algorithm and Gibbs sampler as in the previous sections are used. 

\begin{figure}[H]
\center
\begin{tabular}{cc}
\includegraphics[width = 7.5cm]{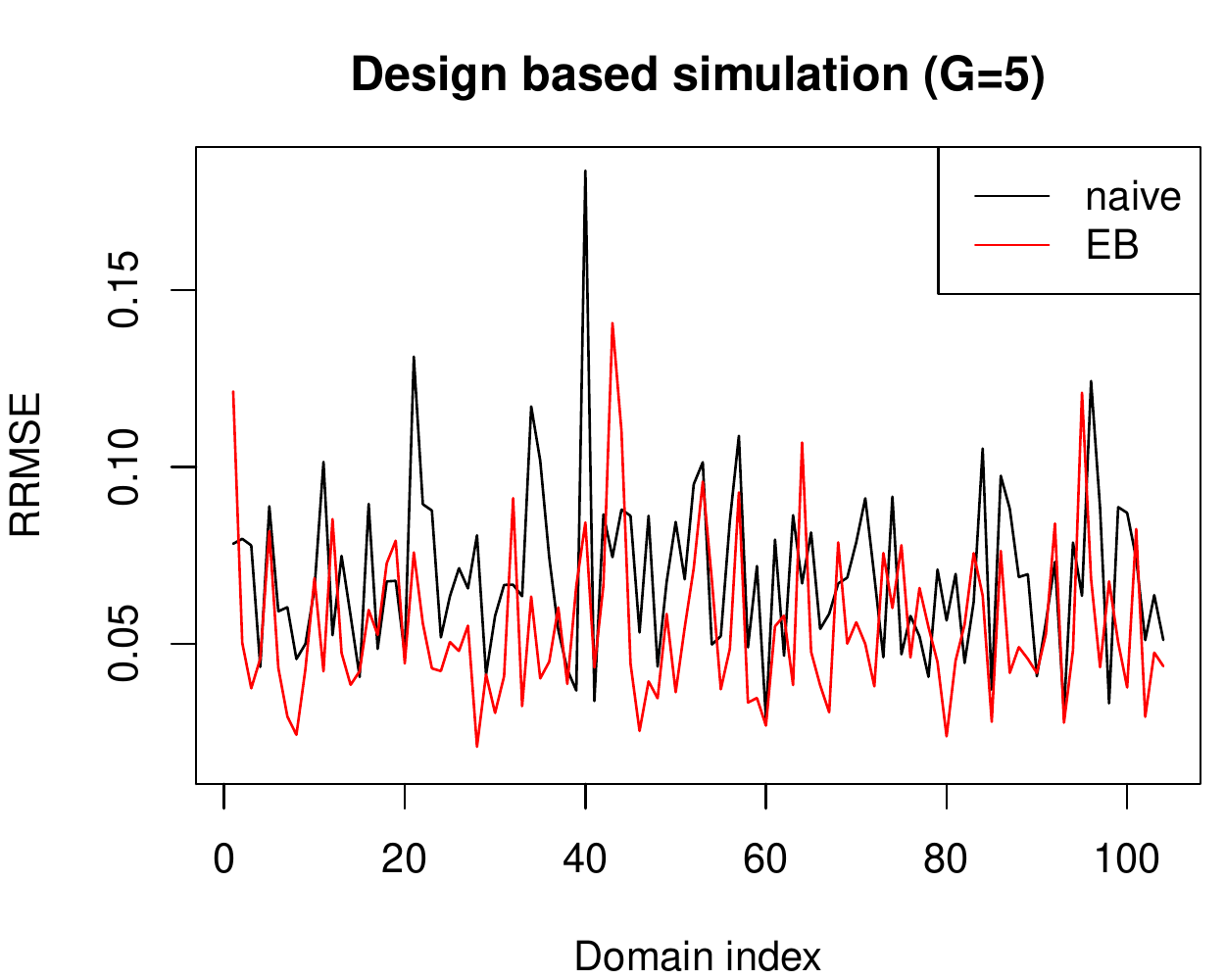}&
\includegraphics[width = 7.5cm]{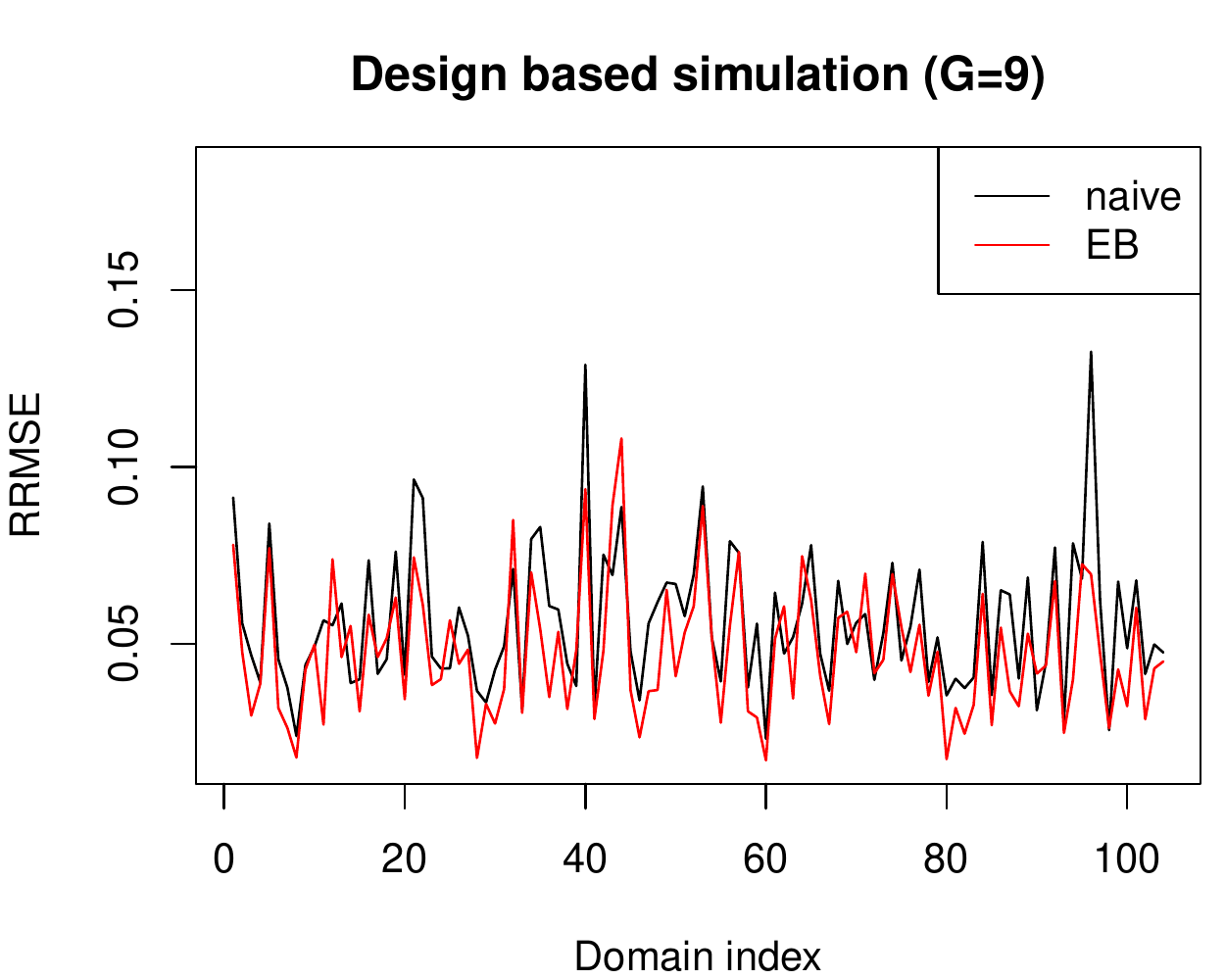}
\end{tabular}
\caption{RRMSE of EB estimator and naive estimator based on design based simulation}
\label{fig:sim2}
\end{figure}

As in the previous sections, the performance of the proposed EB estimator and naive estimator is compared. 
Figure~\ref{fig:sim2} shows the RRMSE for the EB and naive estimators. 
The figure shows that the the EB estimator resulted in the better performance than the naive estimators in terms of RRMSE for most domains. 
In addition, the degree of improvement is larger in the case of $G=5$, where the frequency distributions contain less information. 
Since this simulation setting does not assume a statistical model, we obtained an important implication that the proposed EB estimator performs well even when the statistical model is misspecified. 
This design based simulation can be seen as an empirical evidence to show the usefulness of our proposed method.

\section{Conclusion}
\label{sec:concl}
We have proposed a new model-based small area estimation method for grouped data where only frequency distributions of the quantity of interest are observed at the area-level.
In the proposed model, the observed frequencies are linked with the area-level auxiliary variables through the unit-level latent variables which are modeled in a similar fashion to the nested error regression model.
The model parameter is estimated easily by using the Monte Carlo EM algorithm based on the efficient importance sampling and the EB estimates of small area parameters are calculated by the output of the Gibbs sampler. 
From the application to the real data of Japan and  simulation studies, we have shown that the proposed EB estimator performs better than the naive estimator. 

Because our proposed model is in a general form, it can be applied to a wide variety of datasets. 
However, if we do focus on the income data, especially on the Gini coefficient or other poverty indicators, a probability distribution assumed by the small area model  should provide good fit to the income distribution and provide a straightforward interpretation. 
The present model that assumes the normal distribution after a transformation may be limited in this sense. 
An extension of our model to the parametric income distribution is left for future studies.

\paragraph{ Acknowledgments.}
This work is partially supported by JSPS KAKENHI (\#19K13667, \#18K12754). 
The computational results were obtained by using Ox version 6.21 \citep{D07}.

\appendix
\section{Appendix}
\subsection{Derivation of the full conditional distributions \eqref{eqn:full}}
\label{sec:app1}
Here the full conditional distributions of $\vbt_i$,  $\check{\v}_i$, $\mu_i$ and $\si_i^2$ in \eqref{eqn:full} are derived.
To avoid the notational complexity, we use the notation $p(\cdot)$ as the pdf or pmf for arbitrary random variable.

First, the joint conditional distribution of $\{ \vbt_i, \check{\v}_i, \mu_i, \si_i^2 \}$ given $\y_i$, $p( \vbt_i, \check{\v}_i, \mu_i, \si_i^2 \mid \y_i )$, is given by
\begin{equation*}
p( \vbt_i, \check{\v}_i, \mu_i, \si_i^2 \mid \y_i ) = {p(\y_i, \vbt_i, \check{\v}_i \mid \mu_i, \si_i^2) p(\mu_i) p(\si_i^2) \over p(\y_i) }.
\end{equation*}
Thus it follows that
\begin{equation*}
p( \vbt_i, \check{\v}_i, \mu_i, \si_i^2 \mid \y_i ) \propto p( \y_i, \vbt_i, \check{\v}_i \mid \mu_i, \si_i^2 ) p(\mu_i) p(\si_i^2).
\end{equation*}
Note that $p(\mu_i) = \phi(\mu_i; \x_i^\top\bbeh, \tah^2)$, where $\phi(\cdot; a, b)$ is the pdf of the normal distribution with the mean $a$ and variance $b$ and 
$$
p(\si_i^2) \propto (\si_i^2)^{-(\lah / 2 + 1) - 1} \exp\left( -{\lah \hat{\varphi}_i \over 2\si_i^2} \right).
$$
Because out-of-sample $\check{\v}_i$ is independent of $\{ \y_i, \vbt_i \}$ given $\{ \mu_i, \si_i^2 \}$, it follows that
\begin{align*}
p(\y_i,\vbt_i, \check{\v}_i \mid \mu_i, \si_i^2 ) &= p(\y_i, \vbt_i \mid \mu_i, \si_i^2 ) p(\check{\v}_i \mid \mu_i, \si_i^2), \\
&= p( \y_i \mid \vbt_i, \mu_i, \si_i^2 ) p( \vbt_i \mid \mu_i, \si_i^2 ) p( \check{\v}_i \mid \mu_i, \si_i^2 )
\end{align*}
where
\begin{align*}
p( \vbt_i \mid \mu_i, \si_i^2 ) p( \check{\v}_i \mid \mu_i, \si_i^2 ) = p( \v_i \mid \mu_i, \si_i^2 ) \propto \prod_{j=1}^{N_i}\phi( v_{ij}; \mu_i, \si_i^2 ),
\end{align*}
for $\v_i = ( \vbt_i^\top, \check{\v}_i^\top )^\top = (v_{i1},\dots,v_{in_i}, v_{i,n_i+1},\dots,v_{iN_i})^\top$.
Furthermore, we can write the pmf of $\y_i$ given $\{ \vbt_i, \mu_i, \si_i^2 \}$ as follows:
\begin{align*}
p( \y_i \mid \vbt_i, \mu_i, \si_i^2 ) =& \ \left[ \prod_{j=1}^{\yt_{i1}} I\{ h_\kah(c_0) \leq v_{ij} < h_\kah(c_1) \} \right] \times \left[ \prod_{j=\yt_{i1}+1}^{\yt_{i2}} I\{ h_\kah(c_1) \leq v_{ij} < h_\kah(c_2) \} \right] \\
& \times \cdots \times \left[ \prod_{j=\yt_{i,G-1}+1}^{n_i} I\{ h_\kah(c_{G-1}) \leq v_{ij} < h_\kah(c_G) \} \right],
\end{align*}
where $I\{ \cdot \}$ is the indicator function and $\yt_{ig} = \sum_{g'=1}^g y_{ig'}$, that is, $n_i = \sum_{g'=1}^G y_{ig'}$.
Note that the value of $p(\y_i \mid \vbt_i,\mu_i, \si_i^2)$ only takes 1 or 0.
Hence, the joint conditional distribution of $\{ \vbt_i, \check{\v}_i, \mu_i, \si_i^2 \}$ given $\y_i$ can be written as
\begin{align*}
& p( \vbt_i, \check{\v}_i, \mu_i, \si_i^2 \mid \y_i ) \\
\propto & \ \phi( \mu_i; \x_i^\top\bbeh, \tah^2 ) \times (\si_i^2)^{(-\lah / 2 + 1) - 1} \exp\left( -{\lah \hat{\varphi}_i \over 2\si_i^2} \right) \\
& \times \left[ \prod_{j=1}^{\yt_{i1}}I\{ h_\kah(c_0) \leq v_{ij} < h_\kah(c_1) \} \phi( v_{ij}; \mu_i, \si_i^2 ) \right]  \times \left[ \prod_{j=\yt_{i1}+1}^{\yt_{i2}} I\{ h_\kah(c_1) \leq v_{ij} \leq h_\kah(c_2) \} \phi( v_{ij}; \mu_i, \si_i^2 ) \right] \\
&\times \cdots \times \left[ \prod_{j= \yt_{i,G-1}+1 }^{n_i} I\{ h_\kah(c_{G-1}) \leq v_{ij} < h_\kah(c_G) \} \phi( v_{ij}; \mu_i, \si_i^2 ) \right] \times \prod_{j=n_i+1}^{N_i} \phi( v_{ij}; \mu_i, \si_i^2 ).
\end{align*}
Then, it follows that
\begin{align*}
p( \mu_i \mid \vbt_i, \check{\v}_i, \si_i^2, \y_i ) \propto& \ \phi( \mu_i; \x_i^\top\bbeh, \tah^2 ) \times \prod_{j=1}^{N_i}\phi( v_{ij}; \mu_i, \si_i^2 ) \\
p( \vbt_i \mid \mu_i, \check{\v}_i, \si_i^2, \y_i ) \propto & \  \left[ \prod_{j=1}^{\yt_{i1}}I\{ h_\kah(c_0) \leq v_{ij} \leq h_\kah(c_1) \} \phi( v_{ij}; \mu_i, \si_i^2 ) \right] \\
& \times \left[ \prod_{j=\yt_{i1}+1}^{\yt_{i2}} I\{ h_\kah(c_1) \leq v_{ij} \leq h_\kah(c_2) \} \phi( v_{ij}; \mu_i, \si_i^2 ) \right] \\
&\times \cdots \times \left[ \prod_{j= \yt_{i,G-1}+1 }^{n_i} I\{ h_\kah(c_{G-1}) \leq v_{ij} < h_\kah(c_G) \} \phi( v_{ij}; \mu_i, \si_i^2 ) \right], \\
p( \check{\v}_i \mid \mu_i, \vbt_i, \si_i^2, \y_i ) = & \ \prod_{j=n_i+1}^{N_i}\phi( v_{ij}; \mu_i, \si_i^2 ), \\
p( \si_i^2 \mid \mu_i, \vbt_i, \check{\v}_i, \y_i ) \propto & \ (\si_i^2)^{( -\lah / 2 + 1 ) - 1}\exp \left( -{\lah \hat{\varphi_i} \over 2\si_i^2} \right) \prod_{j=1}^{N_i} \phi( v_{ij}; \mu_i, \si_i^2 ),
\end{align*}
which leads to the full conditional distributions \eqref{eqn:full}.

\subsection{Appendix for the HLS data}\label{sec:app2}
HLS in 2013 was conducted based on the two stage stratified sampling. 
The first stage sampling strata corresponds to the sampling areas used in Population Census in 2010 and the second stage sampling strata consists of the households in the area. 
We have the information which areas are sampled in the first stage and the total number of the households in each area at the time when Population Census in 2010 was conducted. 
We also know which municipality the sampled areas in the first stage belong to.
In the second stage, all households are sampled if the total number of the households in the area is less than 70, otherwise the number of sampled households is approximately 50.
Combining these information,  the sample size in each municipality is estimated.


\end{document}